\documentclass[preprint,sort&compress,12pt,3p]{elsarticle}

\usepackage{moreverb}
\usepackage{amsmath}
\usepackage{algorithmic}
\usepackage{graphics}
\usepackage{graphicx}
\usepackage{subfigure}
\usepackage{enumitem}
\usepackage{color}
\usepackage{framed}
\usepackage{wrapfig}
\usepackage{bm}
\usepackage{mathrsfs}
\usepackage{amsfonts}
\usepackage{multirow}
\usepackage{longtable}
\usepackage{hyperref}
\usepackage{paralist}
\usepackage{indentfirst}
\usepackage{relsize}
\usepackage{extarrows}
\usepackage{lineno,xcolor}
\usepackage{upgreek}
\usepackage{bm}
\usepackage{xcolor}
\usepackage{booktabs}
\usepackage[flushleft]{threeparttable}
\usepackage[linesnumbered,lined,ruled]{algorithm2e}
\usepackage{color, colortbl}
\definecolor{Gray}{gray}{0.9}
\usepackage{geometry}
\newcommand{\eref}[1]{(\ref{#1})}

\hypersetup{
bookmarks=true,
bookmarksopen=true,
bookmarksnumbered=true,
unicode=false,
pdftoolbar=true,
pdfmenubar=true,
pdffitwindow=false,
pdfstartview={FitH},
pdftitle={My title},
pdfauthor={Author},
pdfsubject={Subject},
pdfcreator={Creator},
pdfproducer={Producer},
pdfkeywords={keywords},
pdfnewwindow=true,
colorlinks=true,
linkcolor=blue,
citecolor=blue,
filecolor=magenta,
urlcolor=blue
}

\journal{Computer Methods in Applied Mechanics and Engineering}

\begin{document}
	
    \title{\Large Physics-Informed Multi-LSTM Networks for Metamodeling of Nonlinear Structures}
    
    \author[NU]{Ruiyang~Zhang}
    \author[NU2]{Yang~Liu}
    \author[NU,MIT]{Hao~Sun\corref{cor}}
    \ead{h.sun@northeastern.edu}
    
    \cortext[cor]{Corresponding author. Tel: +1 617-373-3888}
    
    \address[NU]{Department of Civil and Environmental Engineering, Northeastern University, Boston, MA 02115, USA}
    \address[NU2]{Department of Mechanical and Industrial Engineering, Northeastern University, Boston, MA 02115, USA}
    \address[MIT]{Department of Civil and Environmental Engineering, MIT, Cambridge, MA 02139, USA}
    
    \begin{abstract}
    	\small
    	This paper introduces an innovative physics-informed deep learning framework for metamodeling of nonlinear structural systems with scarce data. The basic concept is to incorporate physics knowledge (e.g., laws of physics, scientific principles) into deep long short-term memory (LSTM) networks, which boosts the learning within a feasible solution space. The physics constraints are embedded in the loss function to enforce the model training which can accurately capture latent system nonlinearity even with very limited available training datasets. Specifically for dynamic structures, physical laws of equation of motion, state dependency and hysteretic constitutive relationship are considered to construct the physics loss. In particular, two physics-informed multi-LSTM network architectures are proposed for structural metamodeling. The satisfactory performance of the proposed framework is successfully demonstrated through two illustrative examples (e.g., nonlinear structures subjected to ground motion excitation). It turns out that the embedded physics can alleviate overfitting issues, reduce the need of big training datasets, and improve the robustness of the trained model for more reliable prediction. As a result, the physics-informed deep learning paradigm outperforms classical non-physics-guided data-driven neural networks.
    \end{abstract}
    
    \begin{keyword}
    	\small
    	physics-informed deep learning \sep long short-term memory  \sep metamodeling \sep nonlinear structures \sep LSTM \sep PhyLSTM$^2$ \sep PhyLSTM$^3$
    \end{keyword}
    
    \maketitle

    \section{Introduction}\label{sintro}
    

Numerical simulations are widely utilized for structural analysis and design of complex engineering systems. Many successful computational implementations have been achieved in last several decades for analyzing structural integrity and capacity subjected to dynamic loading. For example, finite element method (FEM) is one of the most popular simulation-based methods for structural dynamic analysis with extensive applications in civil \cite{phuvoravan2005nonlinear,kwasniewski2006finite}, mechanical \cite{smit1998prediction,migliavacca2002mechanical}, and aeronautical engineering \cite{santiuste2010machining,kapidvzic2014finite}. Despite recent advances in computational power (e.g. high-performance computing clusters or facilities), dramatically growing complexity of numerical models still demands prohibitively heavy computation for complex, large engineering problems with nonlinear hysteretic behaviors under dynamic loads. In addition, the computational cost excessively increases especially when numerous simulations are required to account for the optimization \cite{bottasso2014structural,zhang2019cyber} and stochastic uncertainties of external loads (e.g., Monte Carlo simulations \cite{papadrakakis2002reliability,zhang2013advanced,pisaroni2017continuation} or incremental dynamic analysis (IDA) \cite{vamvatsikos2002incremental,vamvatsikos2010incremental,tirca2015improving} of nonlinear structural systems for fragility/reliability analysis).

To address the aforementioned challenge, researchers have explored the use of metamodels to replace the original time-consuming simulation in order to reduce the computational burden. Metamodel is essentially a model of a structure or system model, with parsimonious forms, used to describe the input and output relationship. Traditionally, regression and response surface methodology (RSM) are widely used for metamodeling \cite{durieux2004regression,box1987empirical,khuri2010response} which are based on the polynomial lease-square fitting. These techniques allow fast computation; however, the accuracy is often insufficient for complex systems due to their simplicity and the well-known limitations of using second-order polynomials for approximating highly nonlinear behaviors \cite{simpson2001metamodels}. Kriging \cite{kleijnen2009kriging}, radial basis functions \cite{hussain2002metamodeling}, and polynomial chaos expansions \cite{spiridonakos2015metamodeling} have also been proposed as metamodeling techniques with applications to uncertainty quantification. A review of application of these methods for metamodeling of some engineering systems can be found in \cite{clarke2005analysis}. For the engineering design of dynamic structures and mechanical systems, structural optimization and model updating have been extensively studied and used to simulate structural behaviors \cite{brownjohn2000dynamic,moaveni2009uncertainty,sun2015hybrid}. However, it generally requires excessive computational efforts on calibrating the model especially when the model is of high fidelity with a large number of parameters. To reduce the computational efforts, model order reduction techniques (e.g., proper orthogonal decomposition \cite{du2002model} and equivalent reduction expansion \cite{papadopoulos1996improvement}) have been developed to establish reduced-fidelity metamodels to approximate the high-fidelity models of complex engineering systems \cite{bai2002krylov,guo2019data,zhang2019model}. Nevertheless, the majority of these methods are generally limited to linear or low-order nonlinear systems under stationary conditions, which makes applying these approaches to model highly nonlinear structures intractable.

Recently, artificial neural networks (ANNs) have been proven to be a powerful metamodeling tool and approximator \cite{chen1992neural,tianping1993approximations}, which often outperforms conventional metamodeling techniques in terms of both prediction accuracy and capability of capturing underlying nonlinear input-output relationship for complex systems \cite{fonseca2003simulation}. Researchers have successfully implemented shallow ANNs (e.g., with only a few layers) for metamodeling structural systems under static and dynamic loading during the past decade \cite{ying2009artificial,christiansen2011artificial,lagaros2012neural}. However, due to the simple architecture, shallow ANNs have distinct limitations in modeling time series of complex nonlinear dynamical systems. Thanks to the state-of-the-art advances in artificial intelligence (AI), recent studies have shown that deep learning (e.g., convolutional neural network (CNN) \cite{lecun1995convolutional} and recurrent neural network (RNN) \cite{medsker1999recurrent, mandic2001recurrent}) is a promising approach to establish metamodels for fast prediction of time history response of dynamical systems \cite{lagaros2012neural,wu2018deep,zhang2019deep,zhang2019physics} and material constitutive modeling \cite{wang2018multiscale,wang2019meta}. For example, Zhang et al. \cite{zhang2019deep} successfully developed a deep long short-term memory (LSTM) network for modeling of nonlinear seismic response of structures with large plastic deformation. However, training a reliable deep learning model requires massive (sufficient) data that must contain rich input-output relationship, which typically cannot be satisfied in most engineering problems. Particularly, the ``black-box'' model highly depends on the representative quality of the labeled data that it is fed in, leading to low accuracy and generalizability outside available data (training/validation datasets). Even with rich data, the trained metamodel is uninterpretable and of no physical sense. Furthermore, grand challenges arise when available data is highly incomplete, scarce and/or noisy, e.g., due to (1) ``synthetic'': limited number of computationally intensive simulations of the high-fidelity model for training data generation, or (2) ``sensing'': limited number of recordings, limited number of sensors, low signal-to-noise ratio, and incompleteness of measured state variables. An potential solution to overcome this limitation is to incorporate scientific principles (e.g., partial differential equations, boundary conditions) into deep neural networks to reduce the violation of the embedded physical laws \cite{raissi2018deep,raissi2019physics,sun2019surrogate,yang2019predictive,zhu2019physics,zhang2019physics,kissas2020machine}. To address the aforementioned issues, we develop physics-informed multi-LSTM networks for metamodeling of nonlinear structures and show applications to buildings under earthquake excitation. The key idea is to embed available physics information into deep neural networks, which will boost the learning within a feasible solution space. Such metamodels possess salient features that include (1) clear interpretability with physics meaning, (2) superior generalizability with robust inference, and (3) excellent capability of dealing with less rich data.

This paper is organized as follows. Section \ref{sec:method} introduces two physics-informed multi-LSTM network architectures for structural metamodeling, e.g., the physics-reinforced double-LSTM (e.g., PhyLSTM$^2$) and the physics-reinforced triple-LSTM (e.g., PhyLSTM$^3$). In Section \ref{sec:MRF}, the performance of PhyLSTM$^2$ and PhyLSTM$^3$ is verified through a steel moment-resisting frame with rate-independent hysteresis. Section \ref{sec:bouc-wen} presents another numerical example to compare PhyLSTM$^2$ and PhyLSTM$^3$ for metamodeling of a nonlinear system with rate-dependent hysteresis. Section \ref{sec:conc} summarizes the conclusions. The data and codes used in this paper will be publicly available on GitHub at \href{https://github.com/zhry10/PhyLSTM}{https://github.com/zhry10/PhyLSTM} after the paper is published.

    \section{Physics-informed Multi-LSTM Network for Metamodeling}\label{sec:method}

Metamodeling of structural systems aims to develop reduced-fidelity (or reduced-order) models that effectively capture underlying nonlinear input-output behaviors. A metamodel can be trained on datasets obtained from high-fidelity simulation or actual system sensing. For better illustration, we consider a building-type structure and hypothesize the earthquake dynamics is governed by the reduced-fidelity nonlinear equation of motion (EOM):
    \begin{equation}
        \label{eq:EOM}
        \textbf{M}\ddot{\mathbf{u}} + \underbrace{\textbf{C}\dot{\mathbf{u}} + \lambda \textbf{K}\mathbf{u}+ (1-\lambda)\textbf{K}\mathbf{r}}_{\mathbf{h}} = -\textbf{M}\boldsymbol{\Gamma} {a}_g
        \end{equation}
        where \textbf{M} is the mass matrices; \textbf{C} is the damping matrices; \textbf{K} is the stiffness matrices; \(\mathbf{u}, \dot{\mathbf{u}}\), and \(\ddot{\mathbf{u}}\) are the relative displacement, velocity, and acceleration vector to the ground; \(\mathbf{r}\) is an auxiliary non-observable hysteretic parameter (or called hysteretic displacement); $\lambda \in (0, 1]$ is the ratio of post-yield stiffness to pre-yield (elastic) stiffness; \({a}_g\) represents the ground acceleration; \(\boldsymbol{\Gamma}\) is the force distribution vector; \(\mathbf{h}\) represents the total nonlinear restoring force. The EOM essentially maps the ground motion \(a_g\) to structural response \(\mathbf{u}\), \(\dot{\mathbf{u}}\), \(\ddot{\mathbf{u}}\) and \(\mathbf{r}\). Normalize Eq. \eref{eq:EOM} based on \textbf{M}, the governing equation can be rewritten in a more general form as
        \begin{equation}
        \label{eq:eom}
        \ddot{\mathbf{u}} + \mathbf{g} = -\boldsymbol{\Gamma}{a}_g
    \end{equation}
where $\mathbf{g}(t)=\textbf{M}^{-1}\mathbf{h}(t)$ is the mass-normalized restoring force and $\mathbf{g}(t)=\mathcal{G}\left(\mathbf{Z}(t)\right)$ with \(\mathcal{G}\) being an \textit{unknown latent function}. Here, $\mathbf{Z}$ denotes the state space (SS) variable that includes the displacement $\mathbf{u}$, the velocity $\dot{\mathbf{u}}$, and the hysteretic parameter $\mathbf{r}$, namely, $\mathbf{Z}=\{\mathbf{z}_1,\mathbf{z}_2,\mathbf{z}_3\}^T=\{\mathbf{u},\dot{\mathbf{u}},\mathbf{r}\}^T$. Developing mathematically close form of a nonlinear reduced-fidelity model based on physics (e.g., a parsimonious form of \(\mathbf{g}\)) is intractable especially when the nonlinearity is complex, implicit, and of high order. 

In nonlinear time history analysis of building-type structures under seismic excitation, a fast prediction of the state space variable $\mathbf{Z}$ is of our significant interest. An effective metamodel could establish an efficient and accurate mapping from the seismic input to nonlinear structural response, e.g., $a_g\xrightarrow{\text{metamodel}}\mathbf{Z}$. Our recent study showed that LSTM is a powerful deep learning approach for sequence-to-sequence input-output relationship modeling and thus holds strong promise to serve as a metamodel \cite{zhang2019deep}. However, to train an LSTM-based metamodel, it is essential to have complete state measurement of $\mathbf{Z}$ for a given seismic input $a_g$ (e.g., response data of \(\mathbf{u}\), \(\dot{\mathbf{u}}\) and \(\mathbf{r}\) should be all measured). This is particularly intractable and challenging because the auxiliary hysteretic parameter \(\mathbf{r}\) is typically non-observable and latent which cannot be extracted from large-scale high-fidelity model simulations or from actual system sensing. Yet, predicting such a nonlinear parameter is very important since it reflects the macroscopic nonlinearity of the system (with attributes from local nonlinearity) and relates to the internal hysteretic restoring force. These evidences illustrate that a direct application of a deep learning approach (e.g., LSTM) to establish the metamodel is inapplicable for the above mentioned problem. To address this fundamental challenge, we develop an innovative physics-informed deep learning paradigm (e.g., multi-LSTM networks constrained by physics) for metamodeling of nonlinear structural systems, which systematically maps $a_g$ to the full state $\mathbf{Z}$ given incomplete data (e.g., \(\mathbf{r}\) is not measured). In the following subsections, we introduce the basic concept and algorithm architectures of the proposed new paradigm.



\subsection{LSTM Network}
\begin{figure}[t!]
	\centering
	\subfigure[Deep LSTM network]{\includegraphics[width=0.55\linewidth]{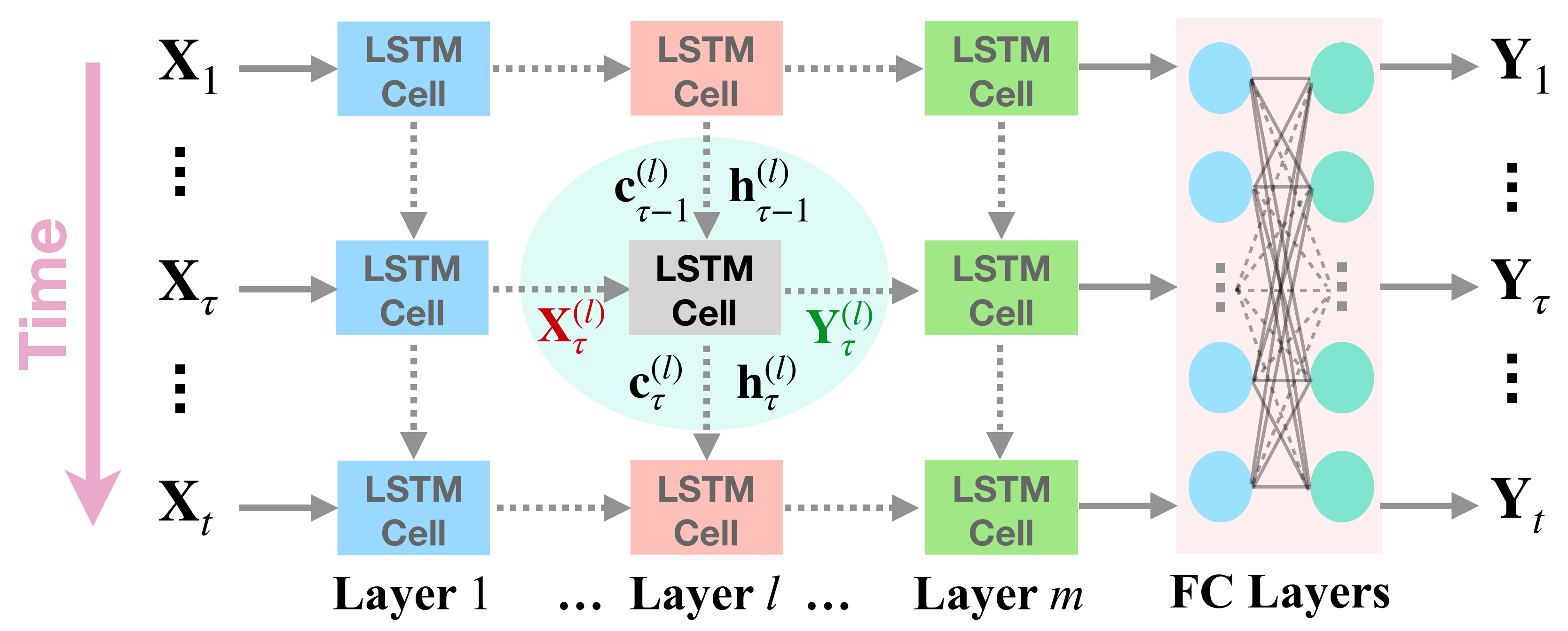} \label{DeepLSTM}} 
	\subfigure[Single LSTM cell structure]{\includegraphics[width=0.42\linewidth]{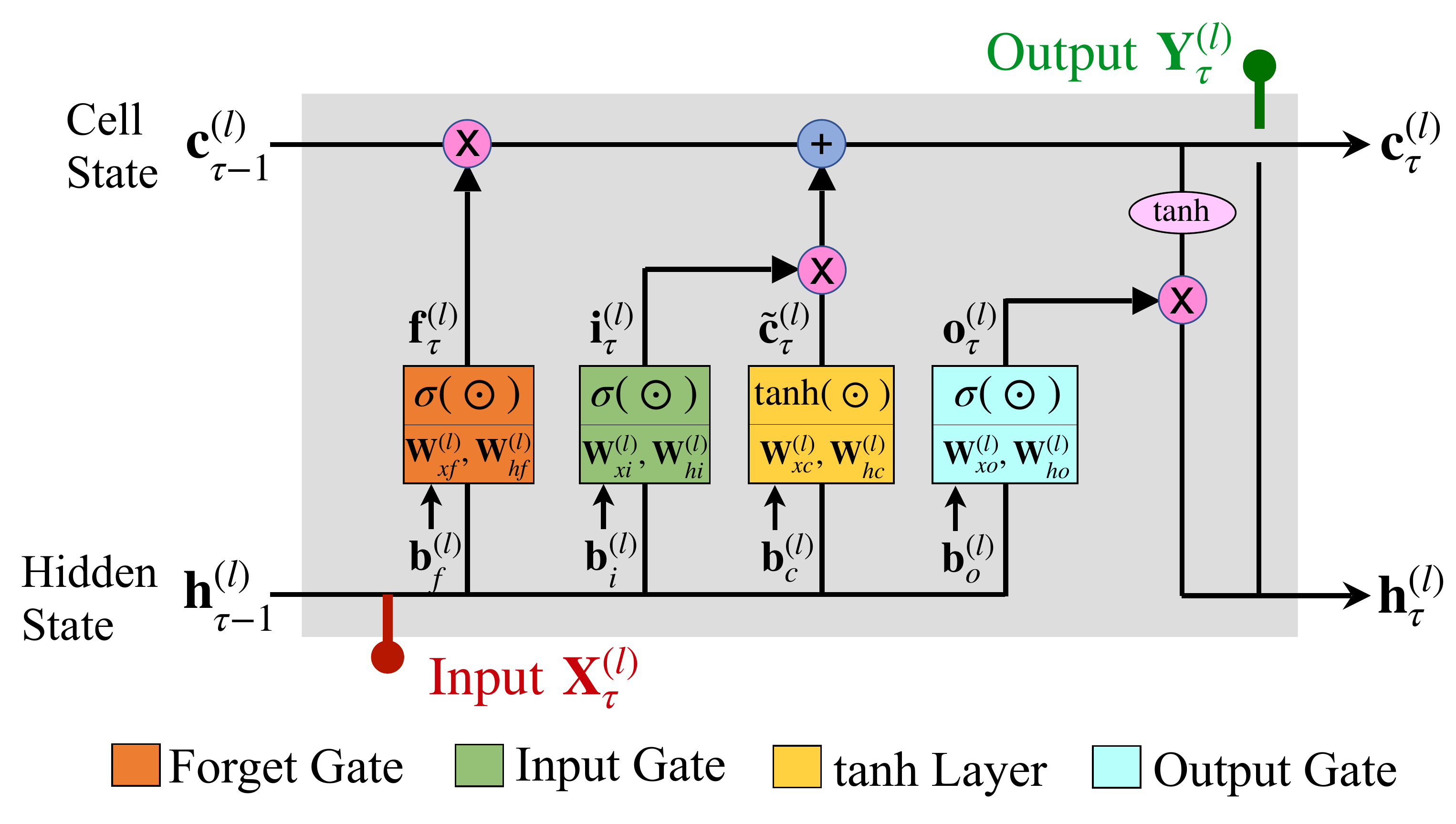} \label{LSTMCell}}
	\caption{Schematic of deep LSTM networks: (a) architecture of a deep LSTM network with $m$ LSTM layers and multiple fully-connected layers for forsequence-to-sequence modeling; (b) architecture of a typical \texttt{LSTM} cell of the $l$th layer at time $t$, which consists of cell input $\mathbf{X}_t^{(l)}$, cell output $\mathbf{Y}_t^{(l)}$, cell state $\mathbf{c}_t^{(l)}$, hidden state $\mathbf{h}_t^{(l)}$, and four gate variables \(\big\{\mathbf{f}_t^{(l)}, \mathbf{i}_t^{(l)}, \tilde{\mathbf{c}}_t^{(l)}, \mathbf{o}_t^{(l)}\big\}\).}
	\label{fig:lstm}
\end{figure}

We first introduce the fundamental algorithm architecture of deep LSTM networks for sequence-to-sequence modeling \cite{zhang2019deep}, which consist of multiple hidden layers (including both LSTM layers and fully connected layers) in addition to the input and output layers as shown in Figure \ref{DeepLSTM}. The deep LSTM network maps the input sequence to the output sequence pairwise in the temporal space ($\tau=1,2,...,t$). To implement the deep LSTM network trained with multiple datasets, both the input and output sequences must be formatted as three-dimensional arrays, where the entries are the samples (e.g., independent datasets) in the first dimension, the time steps in the second dimension, and the input or output features/channels in the third dimension. 

Each LSTM layer contains a suite of LSTM cells as shown in Figure \ref{fig:lstm}. Each LSTM cell, which is very similar to the neural node in classical neural networks, contains an independent set of weights and biases shared across the entire temporal space within the layer. The LSTM cell consists of four interacting units, including an internal cell, an input gate, a forget gate, and an output gate. The internal cell memorizes the cell state at the previous time step through a self-recurrent connection. The input gate controls the flow of input activation into the internal cell state. The output gate regulates the flow of output activation into the LSTM cell output. The forget gate scales the internal cell state, enabling the LSTM cell to forget or reset the cell's memory adaptively. Let us denote, at the time step $t$ ($t=1,...,n$, where $n$ is the total number of time steps) and within the $l$th LSTM network layer, the input state to the LSTM cell as $\mathbf{x}_t^{(l)}$, the forget gate as $\mathbf{f}_t^{(l)}$, the input gate as $\mathbf{i}_t^{(l)}$, the output gate as $\mathbf{o}_t^{(l)}$, the cell state memory  as $\mathbf{c}_t^{(l)}$, and the hidden state output as $\mathbf{h}_t^{(l)}$. At the previous time step $t-1$, we denote the cell state memory as $\mathbf{c}_{t-1}^{(l)}$ and the hidden state output as $\mathbf{h}_{t-1}^{(l)}$. The relationship among these defined variables can be described by the equations as follows (also see Figure \ref{LSTMCell} for schematic illustration): 
\begin{equation}
\label{eq:lstm_f}
\mathbf{f}_t^{(l)} = \sigma\left(\mathbf{W}_{xf}^{(l)} \mathbf{x}_t + \mathbf{W}_{hf}^{(l)} \mathbf{h}_{t-1} + \mathbf{b}_f^{(l)} \right)
\end{equation}
\begin{equation}
\label{eq:lstm_i}
\mathbf{i}_t^{(l)}=\sigma\left(\mathbf{W}_{xi}^{(l)} \mathbf{x}_t + \mathbf{W}_{hi}^{(l)} \mathbf{h}_{t-1} + \mathbf{b}_i^{(l)} \right)
\end{equation}
\begin{equation}
\label{eq:lstm_c_t}
\tilde{\mathbf{c}}_t^{(l)} = \tanh\left(\mathbf{W}_{xc}^{(l)} \mathbf{x}_t + \mathbf{W}_{hc}^{(l)} \mathbf{h}_{t-1} + \mathbf{b}_c^{(l)}\right)
\end{equation}
\begin{equation}
\label{eq:lstm_o}
\mathbf{o}_t^{(l)} = \sigma\left(\mathbf{W}_{xo}^{(l)} \mathbf{x}_t + \mathbf{W}_{ho}^{(l)} \mathbf{h}_{t-1} + \mathbf{b}_o^{(l)}\right)
\end{equation}
\begin{equation}
\label{eq:lstm_c}
\mathbf{c}_t^{(l)} = \mathbf{f}_t^{(l)} \odot \mathbf{c}_{t-1}^{(l)} + \mathbf{i}_t^{(l)} \odot \tilde{\mathbf{c}}_t^{(l)}
\end{equation}
\begin{equation}
\label{eq:lstm_h}
\mathbf{h}_t^{(l)} = \mathbf{o}_t^{(l)} \odot \tanh\left(\mathbf{c}_t^{(l)}\right)
\end{equation}
\noindent where \(\mathbf{W}_{\alpha\beta}^{(l)}\) (with $\alpha=\{x, h\}$ and $\beta=\{f, i, c, o\}$) denotes the weight matrices corresponding to different inputs (e.g., $\mathbf{x}_t^{(l)}$ or $\mathbf{h}_t^{(l)}$) within different gates (e.g., input gate, forget gate, tanh layer or output gate as shown in Figure \ref{LSTMCell}), while \(\mathbf{b}_\beta^{(l)}\) represents the corresponding bias vectors; the superscript $l$ denotes the $l$th layer of the LSTM network. For example, \(\mathbf{W}_{xf}^{(l)}\) and \(\mathbf{W}_{hf}^{(l)}\) are the weight matrices corresponding to input vectors $\mathbf{x}_t$ or $\mathbf{h}_t$, respectively, within the forget gate. Here, $\tilde{\mathbf{c}}_t^{(l)}$ denotes a vector of intermediate candidate values created by a tanh layer shown in Figure \ref{LSTMCell}; \(\sigma\) is the logistic sigmoid function; \(\tanh\) is the hyperbolic tangent function; \(\odot\) denotes the Hadamard product (element-wise product). The complex connection mechanism within each LSTM cell makes the deep LSTM network powerful in sequence modeling, which the fully connected layers are beneficial for mapping the temporal feature maps to the corresponding output space.

\subsection{$\text{PhyLSTM}^2$}

The deep LSTM network introduced in the previous subsection is purely based on data and cannot be used to model latent variables (e.g., $\mathbf{r}$) which are not measured in data. To address this issue, we leverage available physics information (e.g., governing equations, states dependency) and encode it into the network architecture. The basic concept is to use one deep \text{LSTM} network (see Figure \ref{DeepLSTM} \cite{zhang2019deep}) to model the sequence-to-sequence input-output relationship inter-connected, via a graph-based differentiator, with another one/two LSTM network(s) to model the physics. As a result, the multiple connected LSTM networks form a ``one-network'' architecture.

\begin{figure}[t!]
	\centering
	\includegraphics[width=0.95\linewidth]{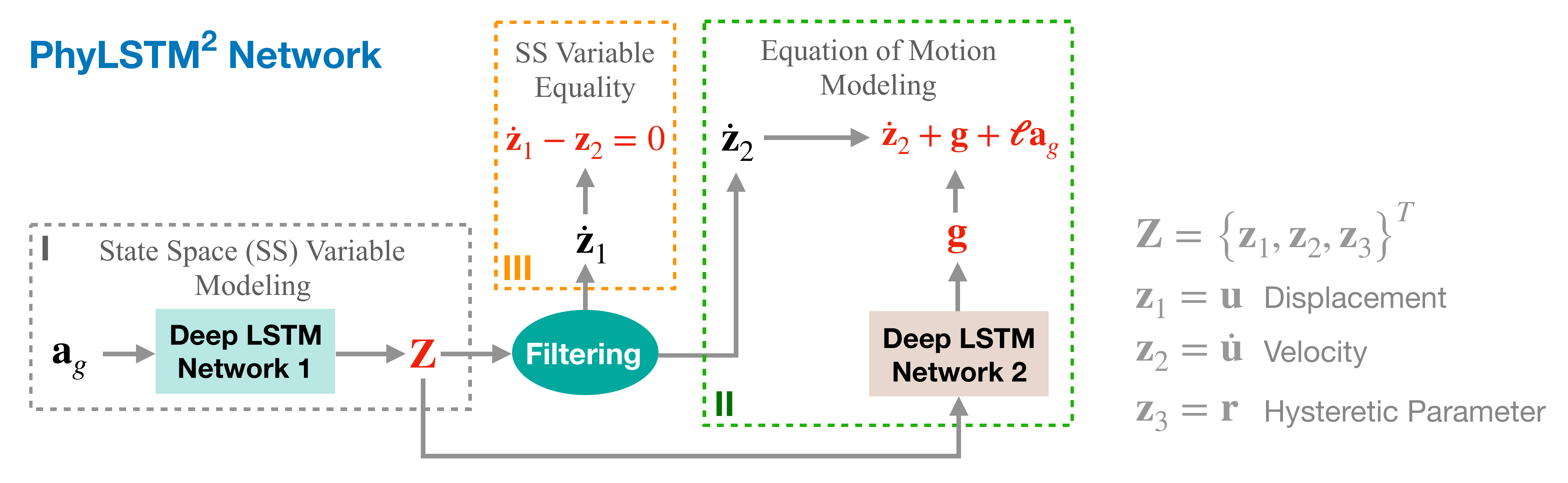}
	\caption{The proposed $\text{PhyLSTM}^2$ network architecture. $\text{PhyLSTM}^2$ consists of two deep LSTM networks for modeling state space variables and nonlinear restoring force. The LSTM networks are interconnected through a graph-based tensor differentiator which calculates the derivative of state space variables.}
	\label{fig:phylstm2}
\end{figure}

Firstly, we introduce the formulation and algorithm architecture of physics-informed double-LSTM network for structural metamodeling ($\text{PhyLSTM}^2$) as shown in Figure \ref{fig:phylstm2}, which consists of three components, including two deep LSTM networks and a graph-based tensor differentiator. To illustrate the concept, we first assemble the structural response to a group of state space variables, \textit{v.i.z.}, \(\mathbf{Z} =\{\mathbf{z}_1, \mathbf{z}_2, \mathbf{z}_3\}^T = \{\mathbf{u}, \dot{\mathbf{u}}, \mathbf{r}\}^T\), each of which has same number of $n$ sample points ranging from $t_1$ to $t_n$, and use one deep LSTM network to establish nonlinear mapping from the ground motion \(a_g\) to the response \(\mathbf{Z}\) (see Box \textbf{\textcolor{black}{\textsf{I}}} in Figure \ref{fig:phylstm2}), e.g., \(\mathbf{Z} = \text{LSTM1}(a_g;\boldsymbol{\theta}_1)\) where \(\boldsymbol{\theta}_1\) denotes the trainable weights and biases of LSTM1. With the available training data \(\{\mathbf{u}_d, \dot{\mathbf{u}}_d\}^T\) (note that $\mathbf{r}$ is an immeasurable latent variable), we can formulate the ``data loss function'' of \text{LSTM1}, written as,
\begin{equation}\label{eq:loss_D}
    \mathcal{J}_d(\boldsymbol{\theta}_1) = \sum_{i=1}^{n_m} \big\|\mathbf{z}_1^{(i)}(\boldsymbol{\theta}_1)-\mathbf{u}_d^{(i)}\big\|_2^2 + \big\|\mathbf{z}_2^{(i)}(\boldsymbol{\theta}_1)-\dot{\mathbf{u}}_d^{(i)}\big\|_2^2
\end{equation}
where $n_m$ is the number of measurement (data) samples. The graph-based differentiation will be realized through finite difference-based filtering, which produces derivatives of $\mathbf{Z}$, namely, \(\dot{\mathbf{Z}} =\{\dot{\mathbf{z}}_1, \dot{\mathbf{z}}_2, \dot{\mathbf{z}}_3\}^T = \{\dot{\mathbf{u}}, \ddot{\mathbf{u}}, \dot{\mathbf{r}}\}^T\). By default, we have the SS variable equality condition $\dot{\mathbf{z}}_1-\mathbf{z}_2\xrightarrow[]{}0$ (see Box \textbf{\textcolor{orange}{\textsf{III}}} in Figure \ref{fig:phylstm2}), leading to the ``equality loss function'':
\begin{equation}\label{eq:loss_e}
\mathcal{J}_e(\boldsymbol{\theta}_1) = \sum_{i=1}^{n_c}\big\|\dot{\mathbf{z}}_1^{(i)}(\boldsymbol{\theta}_1) - \mathbf{z}_2^{(i)}(\boldsymbol{\theta}_1)\big\|_2^2
\end{equation}
where $n_c$ is the number of collocation samples. A second \text{LSTM} network is then used to map the response $\mathbf{Z}$ to the mass-normalized restoring force $\mathbf{g}$ (see Box \textbf{\textcolor{black}{\textsf{II}}} in Figure \ref{fig:phylstm2}), e.g., \(\mathbf{g} = \text{LSTM2}\big(\mathbf{Z}(\boldsymbol{\theta}_1);\boldsymbol{\theta}_2\big)\), where \(\boldsymbol{\theta}_2\) denotes the trainable weights and biases of \text{LSTM2}. Concerning the governing equation in Eq. (\ref{eq:eom}), e.g., $\dot{\mathbf{z}}_2  + \mathbf{g} + \boldsymbol{\Gamma}{a}_g \xrightarrow[]{}0$, we obtain the ``governing loss function'' as 
\begin{equation}\label{eq:loss_g}
\mathcal{J}_g(\boldsymbol{\theta}_1, \boldsymbol{\theta}_2) = \sum_{i=1}^{n_c}\big\| \dot{\mathbf{z}}_2^{(i)}(\boldsymbol{\theta}_1)  + \mathbf{g}^{(i)}(\boldsymbol{\theta}_1, \boldsymbol{\theta}_2) + \boldsymbol{\Gamma}{a}_g \big\|_2^2
\end{equation}
A logical connection of the components in Boxes \textbf{\textcolor{black}{\textsf{I}}}, \textbf{\textcolor{black}{\textsf{II}}} and \textbf{\textcolor{black}{\textsf{III}}} thereby forms the proposed \(\text{PhyLSTM$^2$}\) network, which can be trained by solving the following optimization problem through a standard training algorithm (e.g., gradient descent technique \cite{kingma2014adam}): \begin{equation}
\big\{\hat{\boldsymbol{\theta}}_1, \hat{\boldsymbol{\theta}}_2\big\} = \operatorname*{arg\, min}_{\{\boldsymbol{\theta}_1, \boldsymbol{\theta}_2\}} \mathcal{J}(\boldsymbol{\theta}_1, \boldsymbol{\theta}_2)
\end{equation}
where $\mathcal{J}(\boldsymbol{\theta}_1, \boldsymbol{\theta}_2)$ is the total loss function composed of both data loss and physics loss, given by
\begin{equation}\label{eq:loss}
\mathcal{J}(\boldsymbol{\theta}_1, \boldsymbol{\theta}_2) =  \alpha\mathcal{J}_d(\boldsymbol{\theta}_1) +\beta\mathcal{J}_e(\boldsymbol{\theta}_1) + \gamma\mathcal{J}_g(\boldsymbol{\theta}_1, \boldsymbol{\theta}_2)
\end{equation}
Here, $\alpha$, $\beta$ and $\gamma$ are user-defined weight coefficients for convergence control (e.g., inversely proportional to the magnitude of each term; or for simplicity $\alpha = \beta = \gamma = 1$). The aim here is to optimize the network parameters $\{\boldsymbol{\theta}_1,\boldsymbol{\theta}_2\}$ for both deep LSTM networks such that $\text{PhyLSTM}^2$ can interpret the measurement data while satisfying the physics constraints. Note that the equality condition and the governing equation should hold for any collocation samples that only consist of generic earthquake records with different magnitudes and frequency contents. This will essentially enhance the capability of \text{LSTM1} for modeling the underlying nonlinear input-output relationship within a physically feasible solution space. Note that both LSTM networks in the proposed $\text{PhyLSTM}^2$ architecture used in this study have three LSTM layers and two fully-connected layers.

\begin{figure}[t!]
	\centering
	\includegraphics[width=0.8\linewidth]{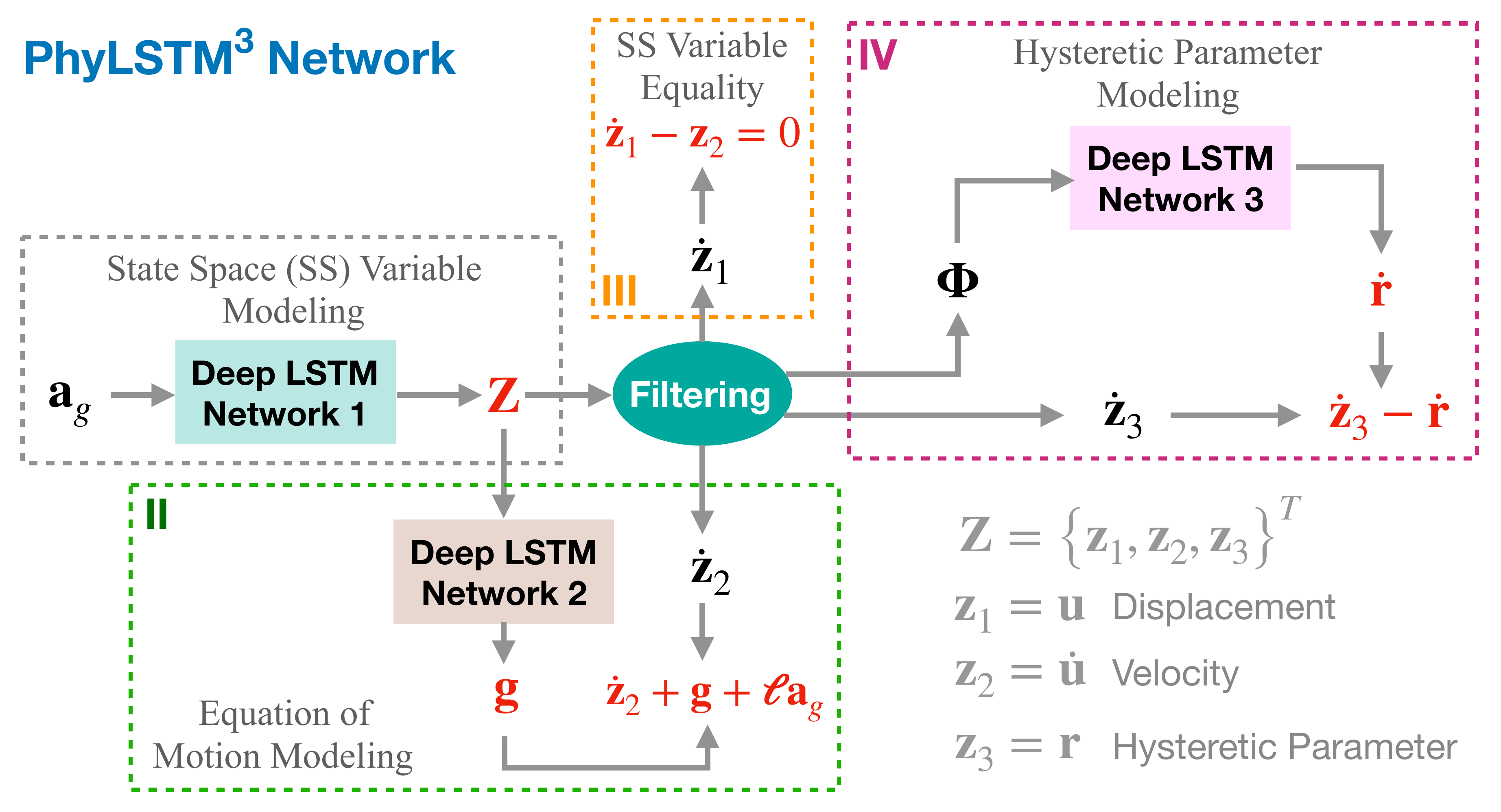}
	\caption{The proposed $\text{PhyLSTM}^3$ network architecture. $\text{PhyLSTM}^3$ network consists of three deep LSTM networks for modeling state space variables, restoring force, and hysteretic parameter. Here,$\boldsymbol{\Phi}$ is a library of system variables, e.g., inspired from the Bouc-Wen model \cite{wen1976method}. The LSTM networks are interconnected through a graph-based tensor differentiator which calculates the derivative of state space variables.}
	\label{fig:phylstm3}
\end{figure}

\subsection{\text{PhyLSTM$^3$}}
For dynamic systems with complex rate-dependent hysteretic behavior (e.g., dependent on \(\dot{\mathbf{r}}\)), the governing equation in Eq. \eref{eq:eom} can be augmented by another nonlinear differential equation of the hysteretic parameter $\mathbf{r}$, expressed as, 
\begin{equation}\label{eq:r_dot}
\vspace{2pt}
    \begin{cases}
        \ddot{\mathbf{u}} + \mathbf{g} = -\boldsymbol{\Gamma}{a}_g\\
        \dot{\mathbf{r}}=f(\boldsymbol{\Phi})\\
    \end{cases}
\vspace{2pt}
\end{equation}
where $f$ is a nonlinear function and $\boldsymbol{\Phi}$ is a library of system variables. For instance, the Bouc-Wen model \cite{wen1976method} takes $\boldsymbol{\Phi} = \big\{\Delta\dot{\mathbf{u}}, |\Delta\dot{\mathbf{u}}|, \mathbf{r}, |\mathbf{r}|^{n-1}, |\mathbf{r}|^n\big\}^T$ to model the nonlinear hysteresis, where $\Delta\dot{\mathbf{u}}$ denotes the inter-story velocity vector. A simplified version of the library reads $\boldsymbol{\Phi} = \left\{\Delta\dot{\mathbf{u}}, \mathbf{r}\right\}^T$ if a priori knowledge is unknown. Therefore, we propose to augment the \text{PhyLSTM$^2$} network by introducing another deep \text{LSTM} network to model the differential equation of \(\mathbf{r}\) (see Box \textbf{\textcolor{magenta}{\textsf{IV}}} in Figure \ref{fig:phylstm3}), e.g., \(\dot{\mathbf{r}} = \text{LSTM3}\big(\boldsymbol{\Phi}(\boldsymbol{\theta}_1);\boldsymbol{\theta}_3\big)\), where \(\boldsymbol{\theta}_3\) denotes the trainable weights and biases of \text{LSTM3}. This essentially forms the \text{PhyLSTM$^3$} network architecture as shown in Figure \ref{fig:phylstm3}, with four components, including three deep LSTM networks and a graph-based tensor differentiator. Similar to $\text{PhyLSTM}^2$, the other two LSTM networks are used to model the state space variables $\mathbf{Z}$ and the mass-normalized restoring force $\mathbf{g}$, respectively. The ``hysteretic loss function'' can then be obtained: 
\begin{equation}\label{eq:h_loss}
\mathcal{J}_h(\boldsymbol{\theta}_1, \boldsymbol{\theta}_3) = \sum_{i=1}^{n_c}\big\| \dot{\mathbf{r}}^{(i)}(\boldsymbol{\theta}_1, \boldsymbol{\theta}_3) - \dot{\mathbf{z}}_3^{(i)}(\boldsymbol{\theta}_1) \big\|_2^2
\end{equation}
The graph-based tensor differentiator calculates the derivative of the state space outputs $\{\dot{\mathbf{z}}_1,\dot{\mathbf{z}}_2,\dot{\mathbf{z}}_3\}$ so that the physics constraints can be well constructed. Note that the \text{PhyLSTM$^3$} network can be trained by optimizing the trainable parameters:
\begin{equation}\label{eq:loss_3lstm}
\big\{\hat{\boldsymbol{\theta}}_1, \hat{\boldsymbol{\theta}}_2, \hat{\boldsymbol{\theta}}_3\big\} = \operatorname*{arg\, min}_{\{\boldsymbol{\theta}_1, \boldsymbol{\theta}_2, \boldsymbol{\theta}_3\}} \big[\mathcal{J}(\boldsymbol{\theta}_1, \boldsymbol{\theta}_2) + \eta\mathcal{J}_h(\boldsymbol{\theta}_1, \boldsymbol{\theta}_3)\big]
\end{equation}
where $\eta$ is also a user-defined weight coefficient (e.g., $\eta = 1$ for simplicity). In \text{PhyLSTM$^3$}, the physics loss enforces the satisfactory of physics constraints including the SS variable equality ($\dot{\mathbf{z}}_1-\mathbf{z}_2\xrightarrow[]{}0$), equation of motion ($\ddot{\mathbf{u}} + \mathbf{g} + \boldsymbol{\Gamma}{a}_g \xrightarrow[]{}0$), and the hysteretic parameter equation ($\dot{\mathbf{z}}_3-\dot{\mathbf{r}}\xrightarrow[]{}0$). Note that \text{PhyLSTM$^3$}, as a generalization of \text{PhyLSTM$^2$}, is, in theory, more powerful in metamodeling of highly nonlinear structures. This will be verified in the numerical example section.

\section{Numerical Validation: 3-story Moment Resisting Frame}\label{sec:MRF}

The proposed physics-informed multi-LSTM networks are firstly validated for metamodeling of a highly nonlinear structural system under seismic excitation. In this example, synthetic data (e.g., nonlinear time-history response) of a 3-story steel moment resisting frame (MRF) are generated by numerical simulation. We test the performance of the proposed \text{PhyLSTM$^2$} and \text{PhyLSTM$^3$} networks for seismic metamodeling of such a structure and compare them with the classical deep LSTM network. Both \text{PhyLSTM$^2$} and \text{PhyLSTM$^3$} map the ground motion $a_g$ to the full state space response $\{\mathbf{u}, \dot{\mathbf{u}}, \mathbf{r}\}^T$ (see Figures \ref{fig:phylstm2} and \ref{fig:phylstm3}), while LSTM can only predict $\{\mathbf{u}, \dot{\mathbf{u}}\}^T$ (see Figure \ref{fig:lstm}(a)), given measured displacements and velocities. Note that, as mentioned previously, the hysteretic parameter $\mathbf{r}$ is a non-observable latent variable. The network training has been performed in the Python environment using TensorFlow \cite{abadi2016tensorflow} which is a popular and well documented open source symbolic math library for machine learning applications developed by Google Brain Team. It offers flexible data flow architecture enabling high-performance training of various types of neural networks on a variety of platforms (CPUs, GPUs, TPUs). Simulations in this paper are performed on a workstation with 28 Intel Core i9-7940X CPUs and 2 NVIDIA GTX 1080Ti GPU cards.



\begin{figure}[t!]
	\centering
	\subfigure[Plan view of the prototype building]{\includegraphics[width=0.39\linewidth]{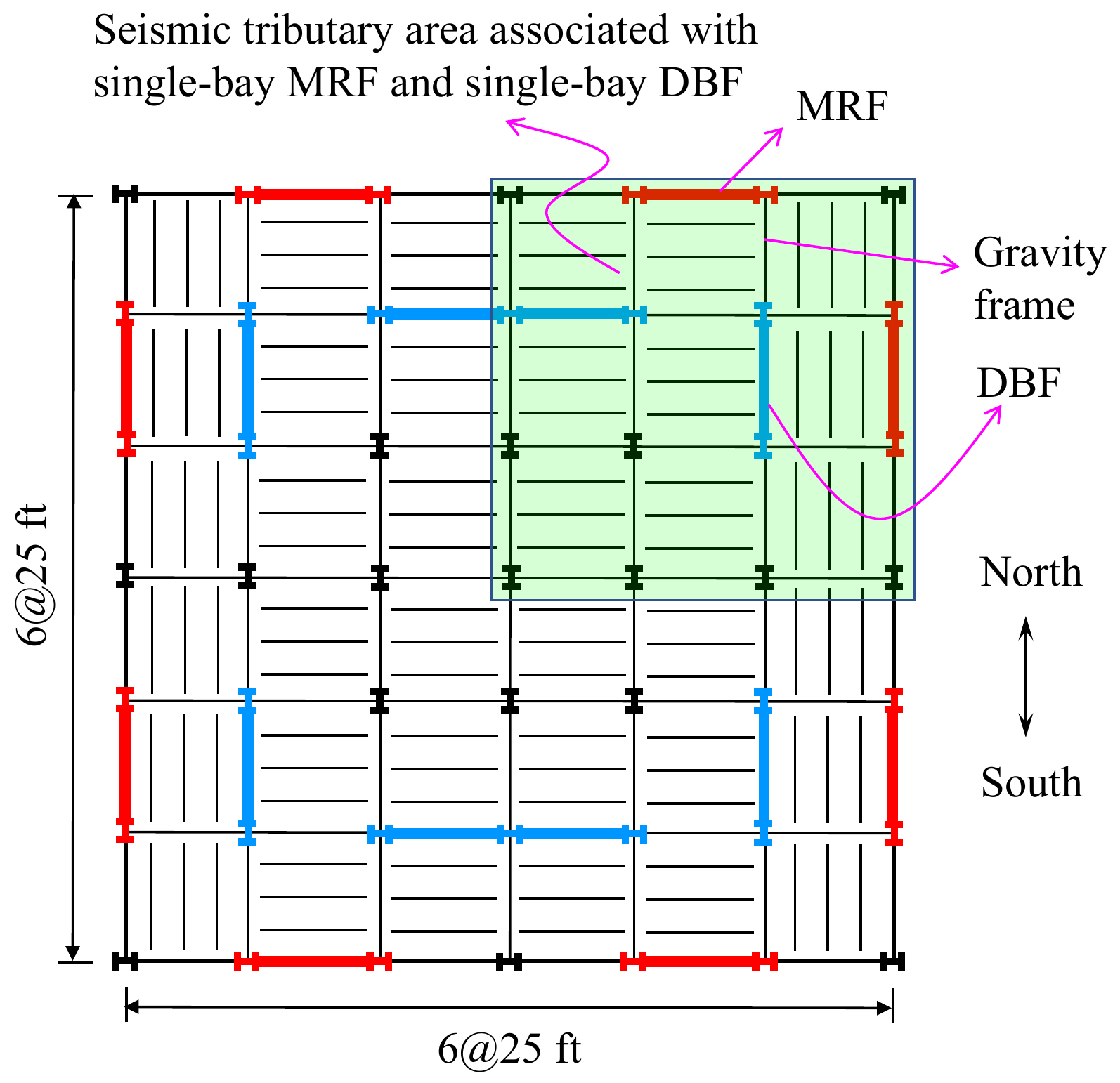}}
	\subfigure[Illustration of the 3-story MRF-DBF structure]{\includegraphics[width=0.59\linewidth]{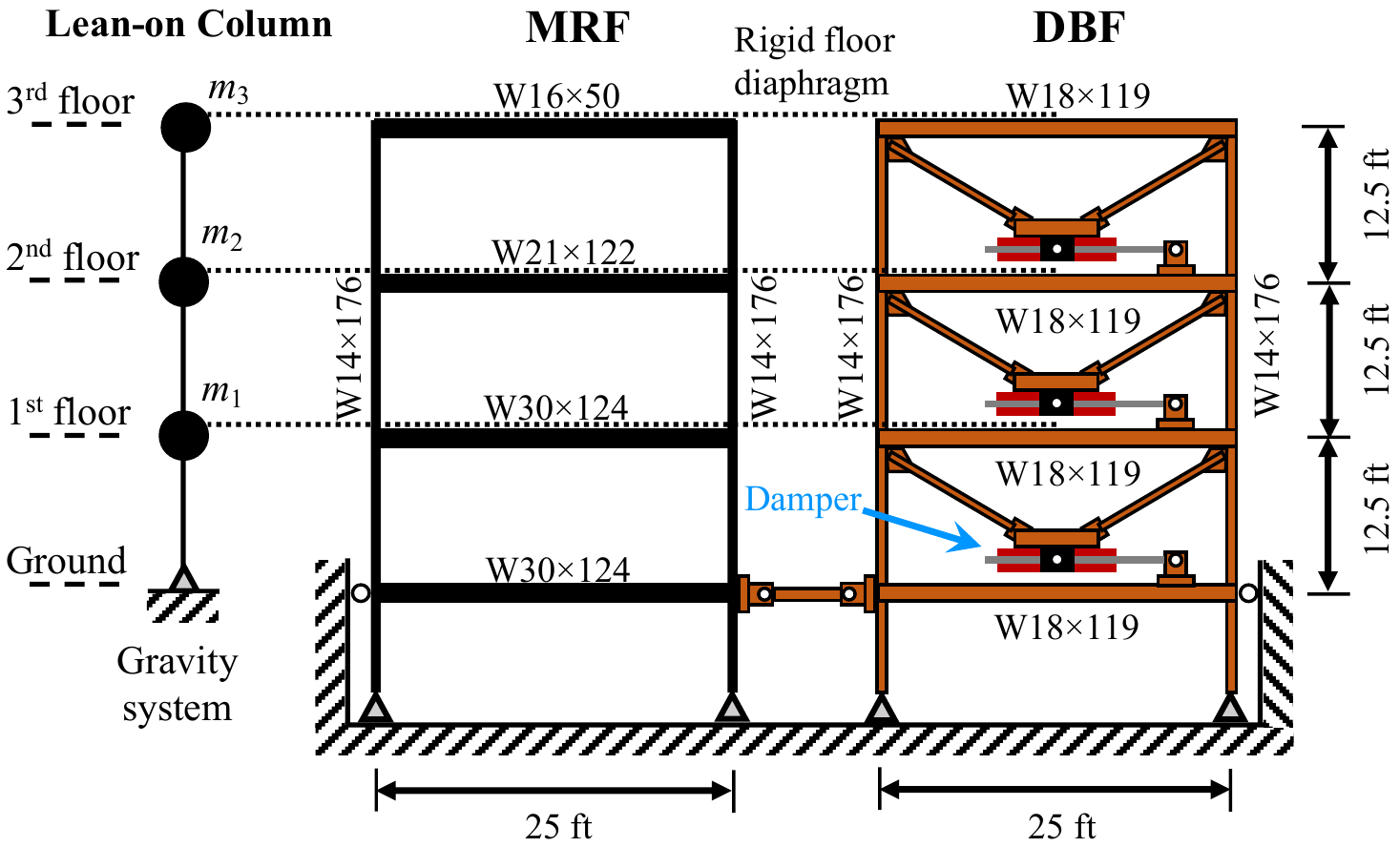}}
	\caption{The 3-story steel MRF building.}
	\label{fig:MRFDBF}
\end{figure}

We test and validate the proposed methodology on a full scale 3-story office building. The prototype building adopted from Dong et al. \cite{dong2018seismic} is assumed to be on a stiff site in Pomona, California. Figure \ref{fig:MRFDBF}(a) shows the plan view of the building. The overall dimensions of the prototype structure are 45.7 m (150 ft) by 45.7 m (150 ft) in plan and 11.43 m (37.5 ft) in elevation. The structural system of the building includes a lateral resisting system, a damping system, and a gravity load system. The lateral resisting system consists of 8 identical single-bay moment resisting frames (MRFs). The damping system consists of 8 single-bay frames with nonlinear viscous dampers and associated bracing, termed as damped braced frames (DBFs). The gravity load system includes the uniformly distributed gravity frames in plan. The floor is assumed to be rigid, and thus the MRFs, DBFs, and the gravity system are assumed to deform together in each horizontal direction. Due to the symmetry of the prototype building, only one quarter of the floor plan within the seismic tributary area as shown in Figure  \ref{fig:MRFDBF}(a) is considered, forming the prototype structure investigated in this study. The 3-story prototype structure shown in Figure \ref{fig:MRFDBF}(b) consists of a single-bay MRF, an associated single-bay DBF, and the gravity load system with associated seismic mass. The horizontal displacement at the ground level is restrained, and the columns are fixed at the base level. The design details of this structure can be found in the reference \cite{dong2018seismic}.

To generate the training/validation datasets, the prototype structure shown in Figure \ref{fig:MRFDBF}(b) is modeled by the nonlinear computational platform, RT-Frame2D, developed in an \textit{embedded} function under the MATLAB/Simulink environment \cite{castaneda2012development, castaneda2013computational}. To preserve stability for nonlinear dynamic analysis, an explicit unconditionally-stable integration scheme is adopted \cite{chen2008development}. A concentrated plasticity model is employed for the nonlinear beam-column elements in RT-Frame2D, assuming that yielding occurs at the element ends. A bilinear moment-curvature hysteresis material model, with kinematic hardening and a post yielding ratio of 2.5\%, is applied. Panel zone elements are used to model the shear deformation and the uniform bending deformation of the MRF panel zones. The element properties include the linear flexural rigidity (EI), axial rigidity (EA), shear rigidity (GA) and yield curvature \(\kappa\). Mass is assigned as \(4.78\times10^5\) kg and \(5.17\times10^5\) kg distributed over beam elements at the first/second and third floor respectively for global mass matrix assembling. The gravity load system is represented by the lean-on column, which is modeled by elastic beam-column elements. The seismic mass is lumped and the gravity load is applied at each floor level on the lean-on column so that P-$\Delta$ effects are included in the nonlinear analysis. The lean-on column is connected to the MRF using a rigid diaphragm. The inherent damping ratios of the first two modes are assigned as 2\% using Rayleigh damping. This does not account for energy dissipation from inelastic response of the MRF, which is included directly within the nonlinear elements. The natural frequencies are 1.02 Hz, 3.61 Hz, and 8.32 Hz for the first three modes. More details of the numerical modeling can be found in \cite{castaneda2012development, castaneda2013computational}. 

\begin{figure}[t!]
	\centering
	\subfigure[Conditional acceleration spectra]{\includegraphics[width=0.485\linewidth]{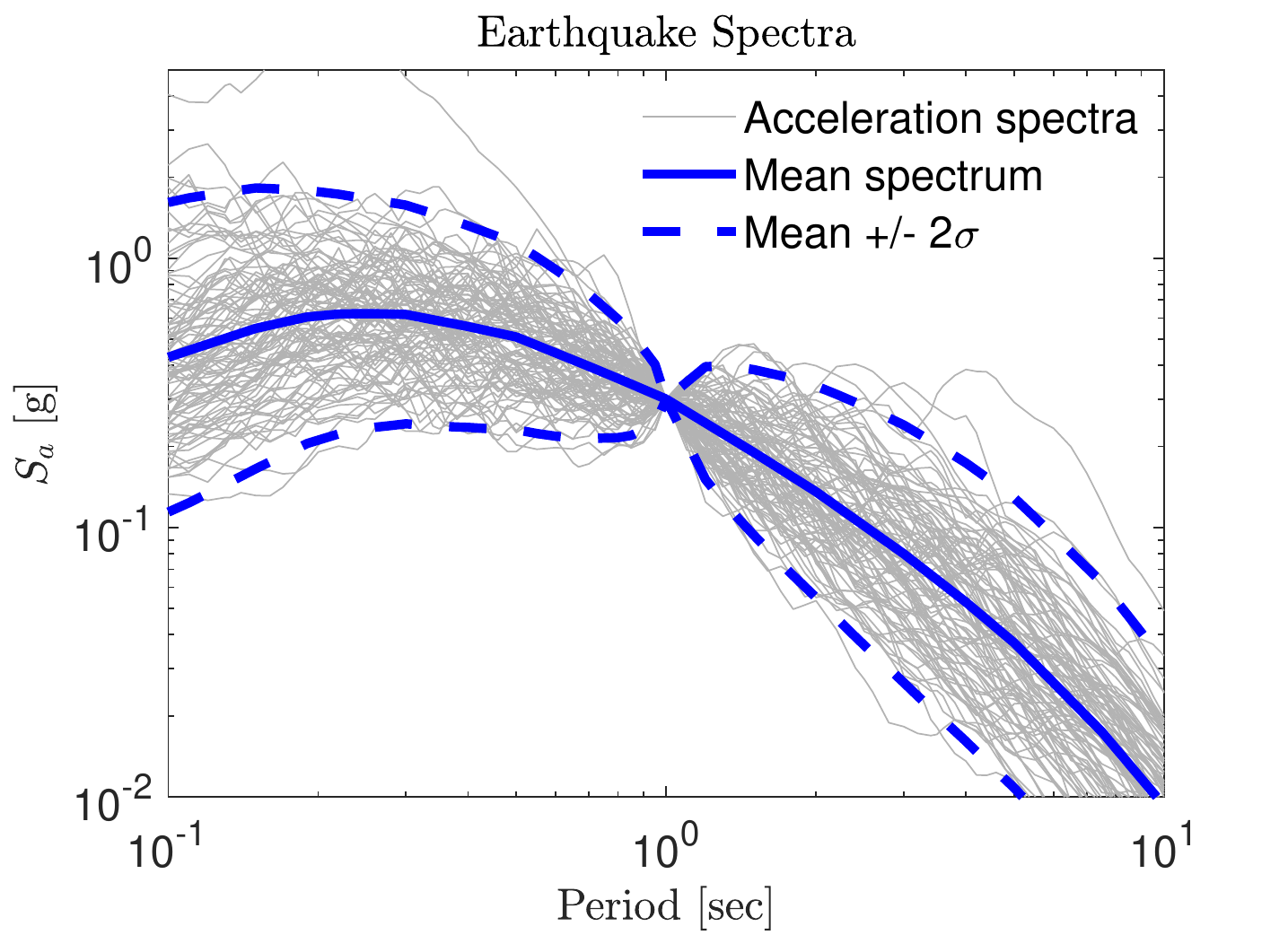}}
	\subfigure[Cluster earthquake spectra centroids]{\includegraphics[width=0.485\linewidth]{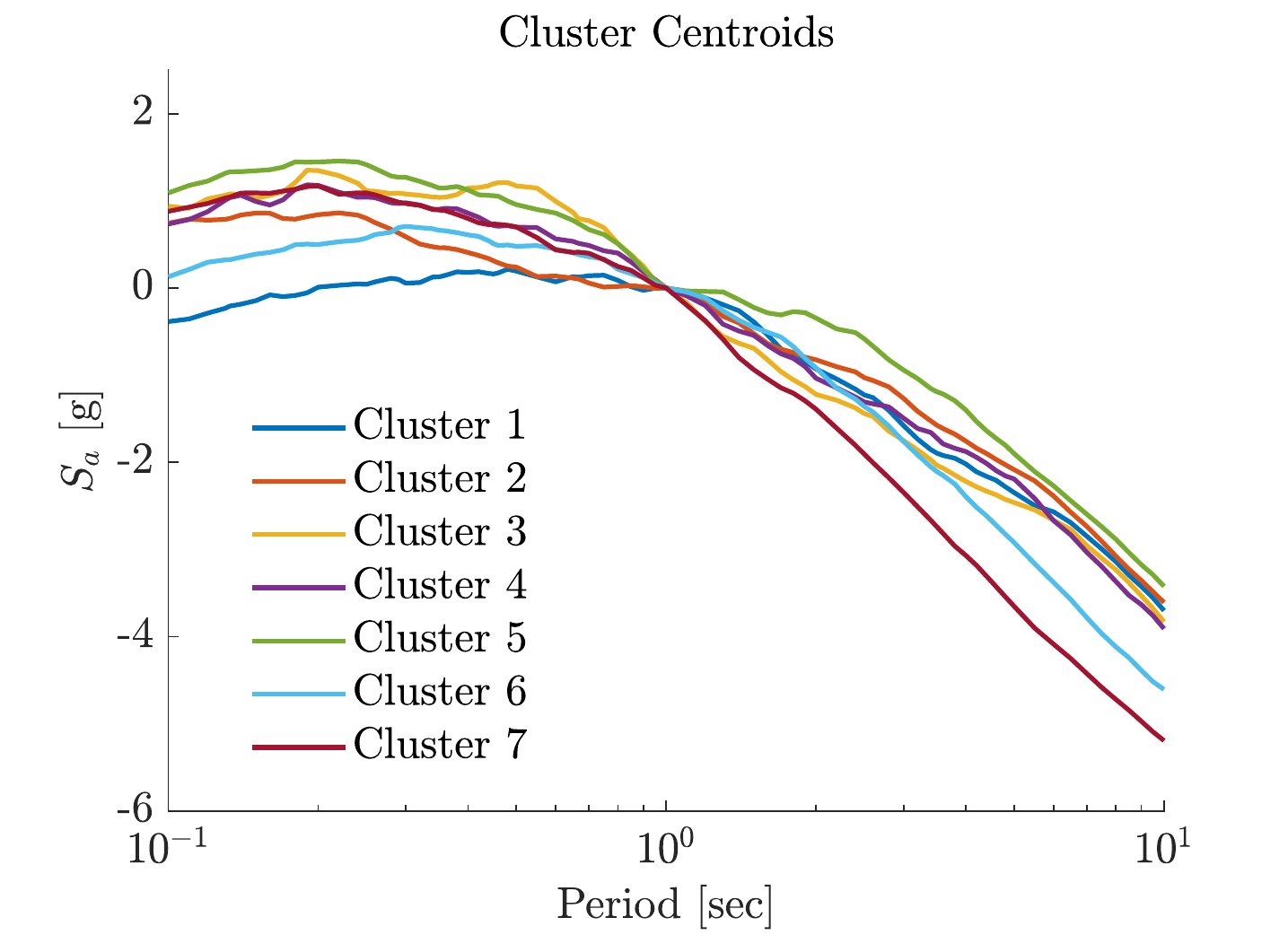}}
	\caption{Suite of earthquake records used in this study.}
	\label{fig:TKmeans}
\end{figure}

A synthetic database, consisting of nonlinear time-history responses of the structure (e.g., \(\{\mathbf{u}_d, \dot{\mathbf{u}}_d\}^T\)), is generated, under excitation of a suite of 97 earthquake records selected from the PEER strong motion database \cite{chiou2008nga} in the area of Pomona, California (latitude, longitude = \(34.0608^\circ\) N, \(117.7558^\circ\) W) with a 10\% probability of exceedance in 50 years. These ground motion records are selected using the earthquake selection and scaling tool developed by Baker and Lee  \cite{baker2018improved} to match the target conditional spectrum which is conditional on a spectral value at a conditioning period of the fundamental natural frequency of the structure. The selected ground motion records are scaled such that the mean response spectrum matches the design spectrum of the prototype building. Figure \ref{fig:TKmeans}(a) shows the conditional acceleration spectra of all 97 selected earthquake records. The incremental dynamic analysis (IDA) is conducted for each ground motion record with scaled intensities to simulate different levels of structural damages and nonlinear responses composed of both elastic and plastic deformation, producing an ensemble of 806 datasets for the prototype structure. Noteworthy, each dataset contains the input ground acceleration and output structural displacements, velocities, and mass-normalized restoring forces (not used in training and only used for testing the predictability of the trained metamodel). 
Since IDA is conducted for magnitude effects, the earthquake excitations are clustered based on the conditional spectral accelerations \((\textit{S}_a)\) shown in Figure \ref{fig:TKmeans}(a). Figure \ref{fig:TKmeans}(b) shows the identified seven cluster centroids for the suite of 97 earthquakes using an unsupervised learning clustering algorithm \cite{zhang2019deep,zhang2019clustering}. Only one earthquake record that is closest to the cluster centroid is selected from each cluster for generating the training/validation datasets, while the rest are considered as the prediction dataset. Therefore, the ground motion selection process, together with IDA, yields only 46 training/validation datasets for 7 selected ground motion records and a total of 760 prediction datasets for the rest 90 earthquakes. It is worth mentioning that both training and validation datasets are considered as ``known'' where \(\{a_g, \mathbf{u}_d, \dot{\mathbf{u}}_d\}^T\) are fully given for training/validating the \text{PhyLSTM$^2$}, \text{PhyLSTM$^3$} and LSTM metamodels, while the prediction dataset is considered as ``unknown ground truth'' only for testing purpose.

All the training/validation datasets are reshaped to 3D arrays in order to be compatible with the data format for LSTM networks, e.g., the input and output sizes are [46, 10001, 1] and [46, 10001, 3]. A ratio of 0.8/0.2 is used for splitting training and validation datasets which are shuffled before each epoch to maximize feature learning from limited data. The datasets are fed into the LSTM network (see Figure \ref{DeepLSTM} or Box \textbf{\textcolor{black}{\textsf{I}}} in Figures \ref{fig:phylstm2} and \ref{fig:phylstm3}) to compute the data loss $\mathcal{J}_d$. A number of 200 earthquake samples in addition to the known earthquake records in the training/validation datasets are used as collocation samples for determining the physics losses (e.g., $\mathcal{J}_e$, $\mathcal{J}_g$, $\mathcal{J}_h$). 
Training the metamodels consists of two phases with different optimization algorithms. In pre-training, Adam (Adamptive Momentum Estimation) is selected as the optimizer with a learning rate of 0.001 and a decay rate of 0.0001 \cite{kingma2014adam} for a total number of $1\times10^4$ epochs. The pre-trained model is further tuned using L-BFGS optimizer which is a quasi-Newton, gradient-based optimization algorithm \cite{liu1989limited}. The network parameters (weights and biases, e.g., $\boldsymbol{\theta}_1$, $\boldsymbol{\theta}_2$ and $\boldsymbol{\theta}_3$) are updated iteratively through back propagation such that the loss function defined in Eq. \eref{eq:loss} or Eq. \eref{eq:loss_3lstm} is minimized. The trained network (e.g., with the minimum validation loss value) is then used as the metamodel to predict structural displacements, velocities, and restoring forces under unknown/unseen ground motions.  

\begin{figure}[t!]
	\centering
	\includegraphics[width=1\linewidth]{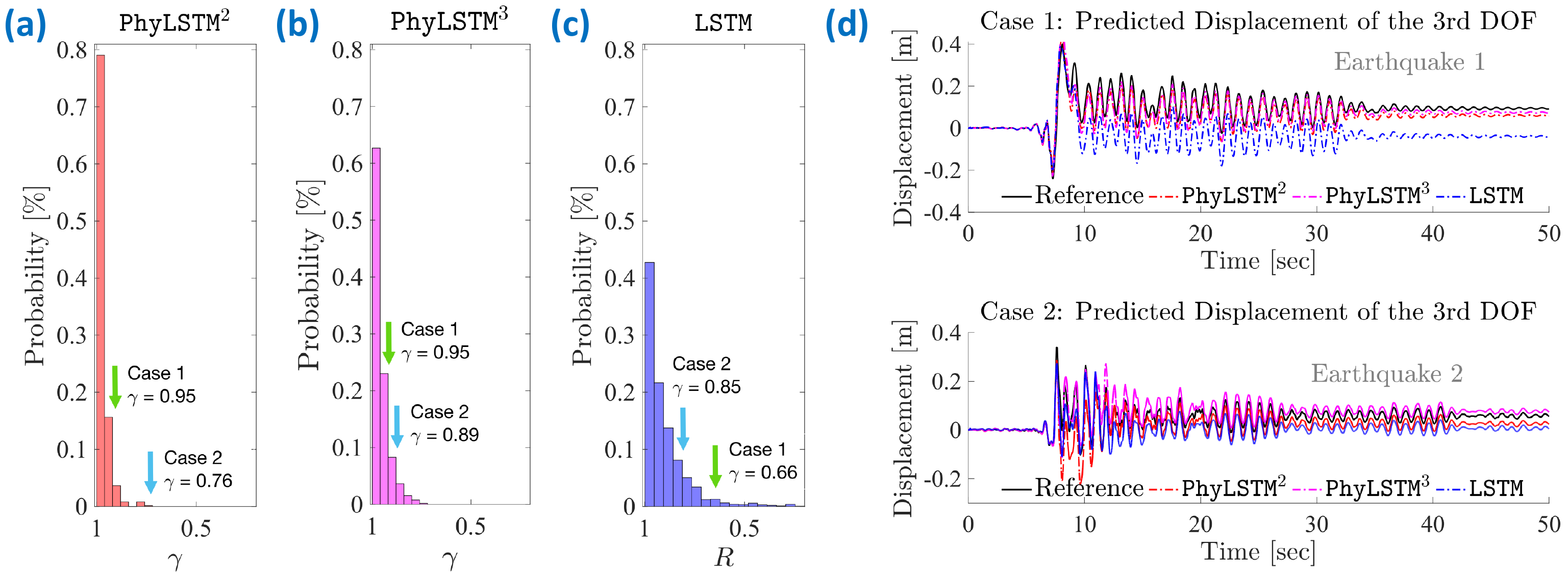}
	\caption{Performance of \text{PhyLSTM$^2$}, \text{PhyLSTM$^3$} and \text{LSTM} for prediction of nonlinear displacements of a 3-story MRF structure: (a)-(c) regression analyses where $\gamma$ denotes the correlation coefficient, and (d) predicted displacements at the top floor under two unseen earthquake excitations randomly picked from the datasets for illustration purpose. Note that Case 1 denotes Earthquake 1 and Case 2 denotes Earthquake 2.}
	\label{fig:mrf_u}
\end{figure}

\begin{figure}[t!]
	\centering
	\includegraphics[width=1\linewidth]{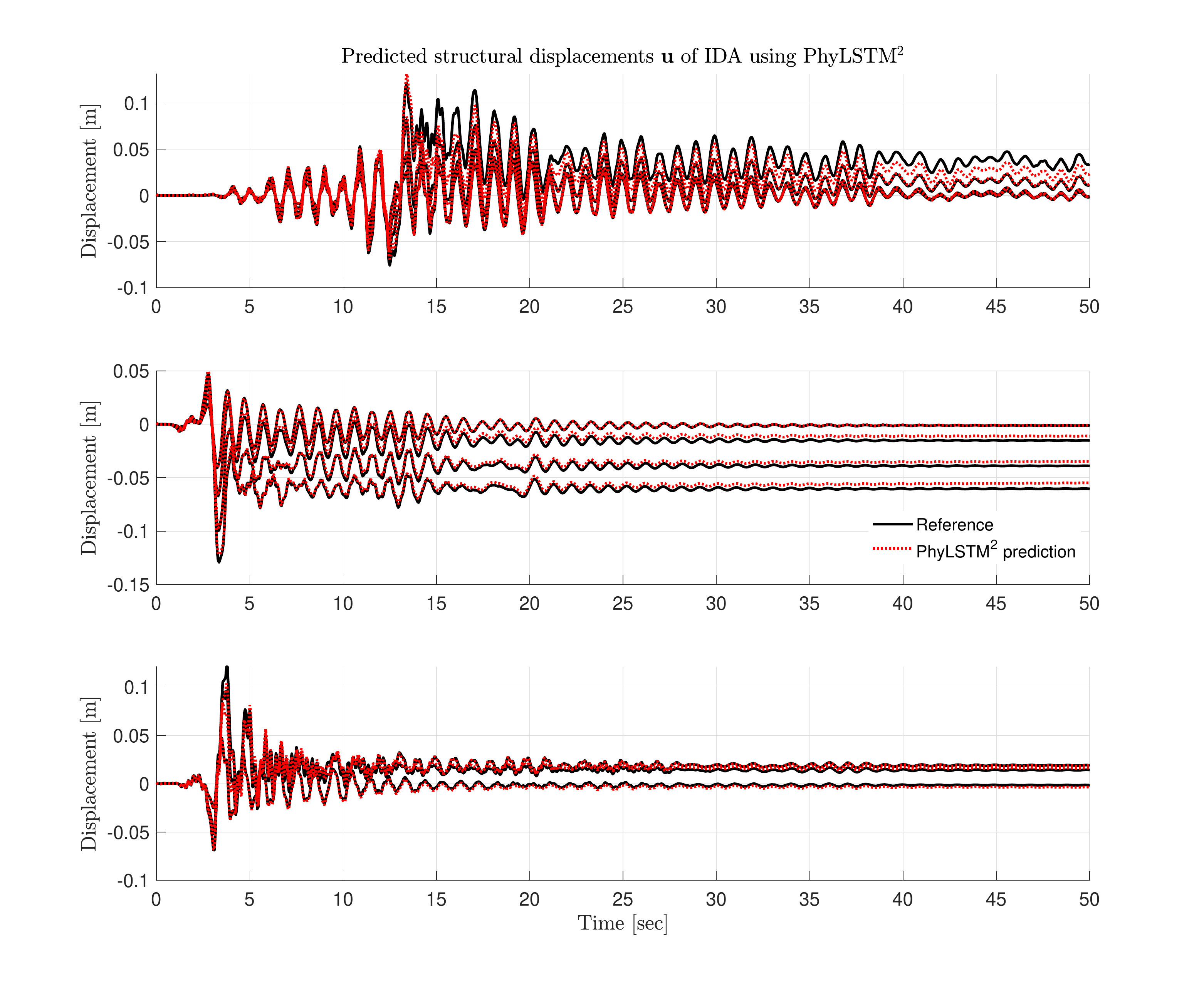}
	\vspace{-36pt}
	\caption{\text{PhyLSTM$^2$}-predicted IDA displacements at the 3rd floor for three example unseen earthquakes with varying intensities (e.g., magnitudes).}
	\label{fig:IDA_2LSTM}
\end{figure}

Figure \ref{fig:mrf_u} shows the performance of the three networks (e.g., $\text{PhyLSTM}^2$, $\text{PhyLSTM}^3$ and LSTM) for prediction of nonlinear displacements of the 3-story MRF structure. Figure \ref{fig:mrf_u}(a)-(c) summarize regression analysis of the predicted displacement time histories across all 760 testing datasets. It can be observed that the majority of the correlation coefficients (denoted as $\gamma$) for both $\text{PhyLSTM}^2$ and $\text{PhyLSTM}^3$ are greater than 0.9, indicating very accurate prediction. Clearly, the proposed physics-informed multi-LSTM approaches are much more robust and produce more accurate prediction compared to classical LSTM without embedded physics. The worst scenario for LSTM corresponds the correlation coefficient $\gamma=0.25$ which is much lower compared to $\text{PhyLSTM}^2$ with $\gamma=0.74$ and $\text{PhyLSTM}^3$ with $\gamma=0.76$. Figure \ref{fig:mrf_u}(d) shows predicted displacement time histories at the top floor under two example earthquakes, with the corresponding correlation coefficients marked in the regression plot for $\text{PhyLSTM}^2$ ($\gamma$ = 0.95 and 0.76), $\text{PhyLSTM}^3$ ($\gamma$ = 0.95 and 0.89), and LSTM ($\gamma$ = 0.66 and 0.85). The $\text{PhyLSTM}^2$ prediction, with $\gamma = 0.95$, matches the reference well in magnitudes, phases, as well as residual drifts that reflect plastic deformation as shown in Figure \ref{fig:mrf_u}(d). Note that the prediction displacement time histories for $\gamma>0.95$ are not shown since the predicted displacements have an excellent match with the ground truth. Even for the case with less satisfactory prediction (e.g., $\gamma=0.76$), the $\text{PhyLSTM}^2$ approach is still able to reasonably well predict the displacement time histories using very limited training data. Similar prediction performance is observed for the $\text{PhyLSTM}^3$ metamodel. The predicted structural displacements using LSTM are also presented in Figure \ref{fig:mrf_u}(d). Although the predicted peak magnitudes and phases of displacements relatively well match the reference, the residual drifts (e.g., plastic deformation) cannot be accurately predicted by LSTM. This indicates that it is intractable to learn the complex hysteretic behavior purely from data in training especially when available datasets are limited. In summary, both \text{PhyLSTM$^2$} and \text{PhyLSTM$^3$} outperform \text{LSTM}, while \text{PhyLSTM$^2$} produces slightly better prediction compared with \text{PhyLSTM$^3$}. Note that the nonlinear hysteresis of this structure is rate-independent (e.g., independent on \(\dot{\mathbf{r}}\)) such that \text{PhyLSTM$^2$} is more capable of modeling the latent nonlinearity given its parsimonious architecture compared with \text{PhyLSTM$^3$}.
The favorable performance of \text{PhyLSTM$^2$}, for example, is further illustrated in Figure \ref{fig:IDA_2LSTM}, which shows the predicted IDA displacements in comparison with the ground truth under excitation of the same earthquake but with varying intensities. It is seen that, although the input earthquakes are scaled linearly, the trained metamodel is capable of capturing and distinguishing the nonlinear structural responses.

\begin{figure}[t!]
	\centering
	\includegraphics[width=1\linewidth]{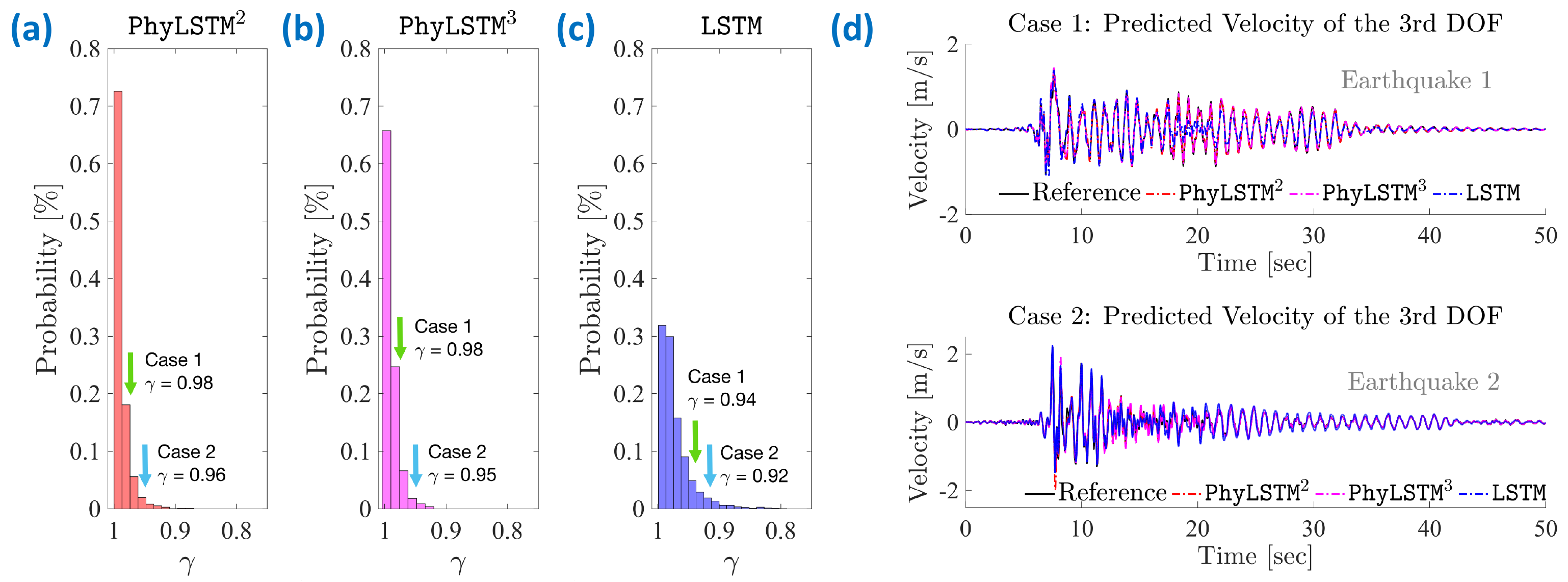}
	\caption{Performance of \text{PhyLSTM$^2$}, \text{PhyLSTM$^3$} and \text{LSTM} for prediction of velocities of a 3-story MRF structure: (a)-(c) regression analyses where $\gamma$ denotes the correlation coefficient, and (d) predicted velocities at the top floor under two unseen earthquake excitations randomly picked from the datasets for illustration purpose. Note that Case 1 denotes Earthquake 1 and Case 2 represents Earthquake 2.}
	\label{fig:mrf_ut}
\end{figure}

\begin{figure}[t!]
	\centering
	\includegraphics[width=1\linewidth]{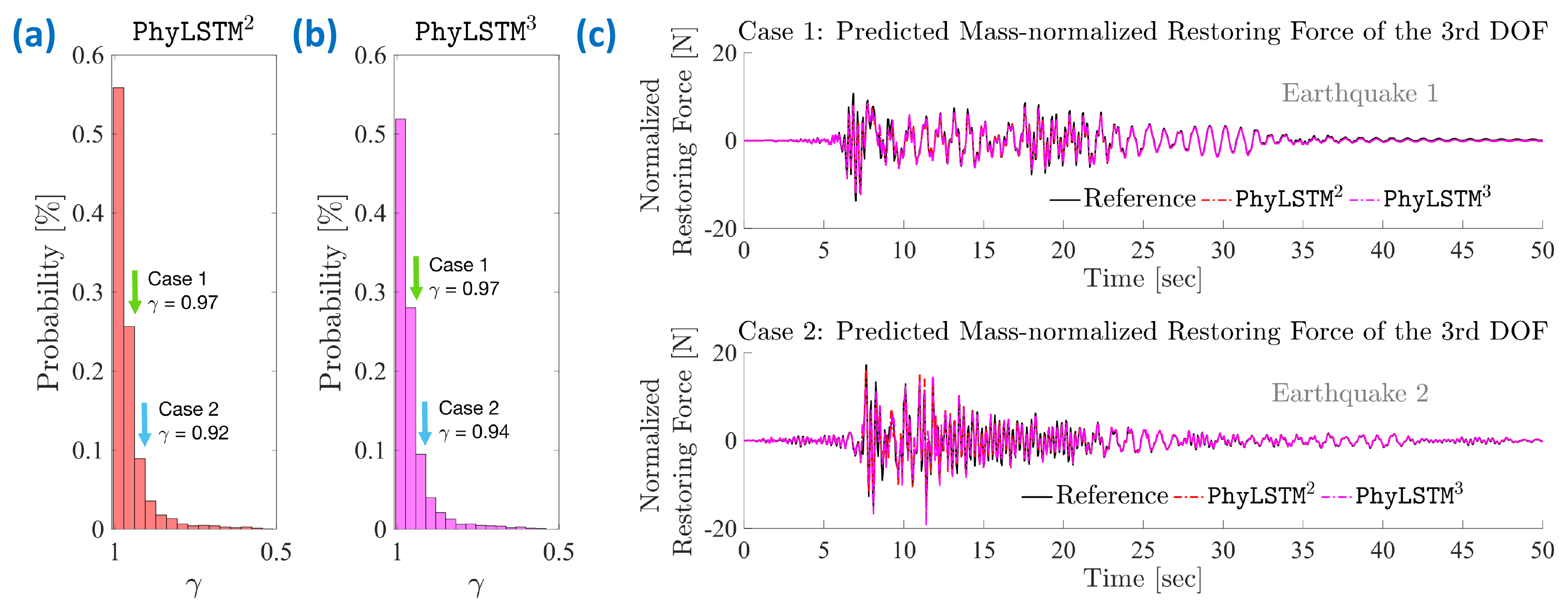}
	\caption{Performance of \text{PhyLSTM$^2$} and \text{PhyLSTM$^3$} for prediction of the mass-normalized restoring forces $\mathbf{g}$: (a)-(b) regression analyses where $\gamma$ denotes the correlation coefficient, and (d) predicted mass-normalized restoring forces at the top floor under two unseen earthquake excitations randomly picked from the datasets for illustration purpose. Note that without the measurements of $\mathbf{g}$, the physics-informed multi-LSTM approaches are able to predict the latent nonlinear restoring force while LSTM fails to predict it without measurement in training.}
	\label{fig:mrf_g}
\end{figure}

Figure \ref{fig:mrf_ut} presents the result of predicted velocities by $\text{PhyLSTM}^2$, $\text{PhyLSTM}^3$ and LSTM, respectively. It turns out the velocities are much easier to learn and can be accurately predicted even using LSTM, because velocity time histories have less complex behaviors such as residuals. Nevertheless, $\text{PhyLSTM}^2$ and $\text{PhyLSTM}^3$ still provide better prediction accuracy compared with the data-driven LSTM. Another advantage of physics-informed multi-LSTM networks is that the latent state (e.g., the hysteretic parameter \(\mathbf{r}\) resulting from LSTM1 or the nonlinear restoring force \(\mathbf{g}\) from LSTM2, as shown in Figure \ref{fig:phylstm2} and \ref{fig:phylstm3}) can be predicted even though no measurement of the state is available for training. This can be realized by the physical knowledge encoded in the network. For example, Figure \ref{fig:mrf_g} shows the predicted mass-normalized restoring force using $\text{PhyLSTM}^2$ and $\text{PhyLSTM}^3$ given no measurements of which in training. This is a mission impossible by classical data-driven LSTM networks. Note that the time history examples shown in Figures \ref{fig:mrf_u}, \ref{fig:mrf_ut}, and \ref{fig:mrf_g} are subjected to the same set of ground motion excitations for better comparison. This example clearly illustrates the accuracy and robustness of the proposed physics-informed multi-LSTM metamodels compared with the classical data-driven LSTM. From the aforementioned results, we can also conclude that, with physics constraints, the proposed physics-informed multi-LSTM metamodels are capable of learning and recognizing hidden patterns obeying given governing laws from very limited data.

\section{Numerical Validation: Bouc-Wen Hysteresis Model}\label{sec:bouc-wen}
We herein consider a nonlinear system with rate-dependent hysteresis (e.g., dependent on \(\dot{\mathbf{r}}\)) as described in Eq. \eref{eq:r_dot} and compare the capability of $\text{PhyLSTM}^2$ and $\text{PhyLSTM}^3$ for complex hysteresis modeling. 
The Bouc-Wen model \cite{wen1976method, zhang2017shake} is adopted for showcase, in which, for the $i$th degree-of-freedom (DOF), the rate-dependent hysteresis is expressed as  \cite{sato1998adaptive}:
\begin{equation}
\label{eq:BoucWen}
\dot{r}_i = \Delta \dot{u_i} - \alpha_i |\Delta\dot{u}_i| |r_i|^{n_i-1}r_i - \beta_i \Delta\dot{u}_i |r_i|^{n_i}
\end{equation}
where $\Delta\dot{u}_i$ is the relative velocity between $(i-1)$th and $i$th DOF, denoted as $\Delta\dot{u}_i=\dot{u}_i-\dot{u}_{i-1}$ for $i\geq2$ and $\Delta\dot{u}_i=\dot{u}_1$ if $i=1$; \(\alpha_i\),  \(\beta_i\) and \(n_i\) are the nonlinear parameters of the Bouc-Wen model. In this example, a single DOF (SDOF) Bouc-Wen model is used with the following parameters: $m=500$ kg, $c=0.35$ kNs/m, $k=25$ kN/m, $\alpha=2$,  $\beta=2$ and $n=3$. The natural frequency of the system is 1.13 Hz. The parameter $\lambda$ in Eq. \eref{eq:EOM} is assumed as 0.5. 
A synthetic database, consisting of 100 samples (e.g., independent seismic sequences), was generated by numerical simulation for the SDOF nonlinear system excited by random band-limited white noise (BLWN) ground motions with different magnitudes. Each simulation was executed up to 30 seconds with a sampling frequency of 50 Hz resulting in 1501 data points for each record. All datasets are formatted to required 3D arrays for $\text{PhyLSTM}^2$ and $\text{PhyLSTM}^3$. Only 10 datasets with BLWN input and corresponding structural displacement and velocity responses are randomly selected and considered as ``known'' datasets for training/validation (with a split ratio of 0.8/0.2), while the rest are considered as ``unknown'' datasets to test the prediction performance of trained metamodels. 50 additional collocation samples (e.g., BLWN input records only) are used to guide the model training with physics constraints.

The network configuration for this example is given as follows: each LSTM network in $\text{PhyLSTM}^2$ and $\text{PhyLSTM}^3$ has two LSTM layers and one FC layer, which turns out to be sufficient to train an accurate model. 
The $\text{PhyLSTM}^2$ and $\text{PhyLSTM}^3$ models are first pre-trained using the Adam optimizer \cite{kingma2014adam} with a learning rate of 0.001 for 5000 epochs and with a learning rate of 0.0001 for another 5000 epochs. Then the L-BFGS optimizer \cite{liu1989limited} is used to enhance the pre-trained model until the default convergence criteria is triggered. We take $\boldsymbol{\Phi} = \{\Delta\dot{u}, r\}^T$ as the simplified library of basis functions for hysteresis modeling.

\begin{figure}[t!]
	\centering
	\includegraphics[width=1\linewidth]{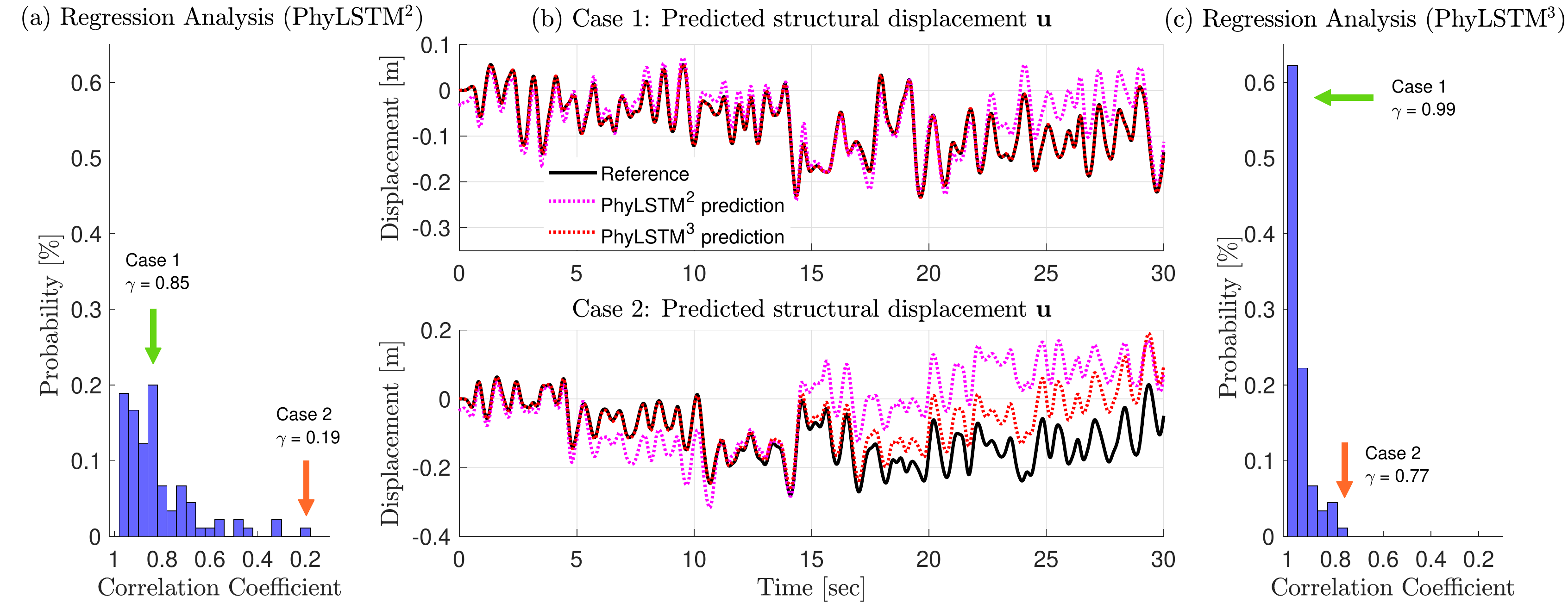}
	\caption{Prediction performance of displacement $u$ using $\text{PhyLSTM}^2$ and $\text{PhyLSTM}^3$: (a) regression analysis for $\text{PhyLSTM}^2$; (b) two examples of predicted displacement time histories; and (c) regression analysis for $\text{PhyLSTM}^3$.}
	\label{fig:u_2LSTM_3LSTM}
\end{figure}

\begin{figure}[t!]
	\centering
	\includegraphics[width=1\linewidth]{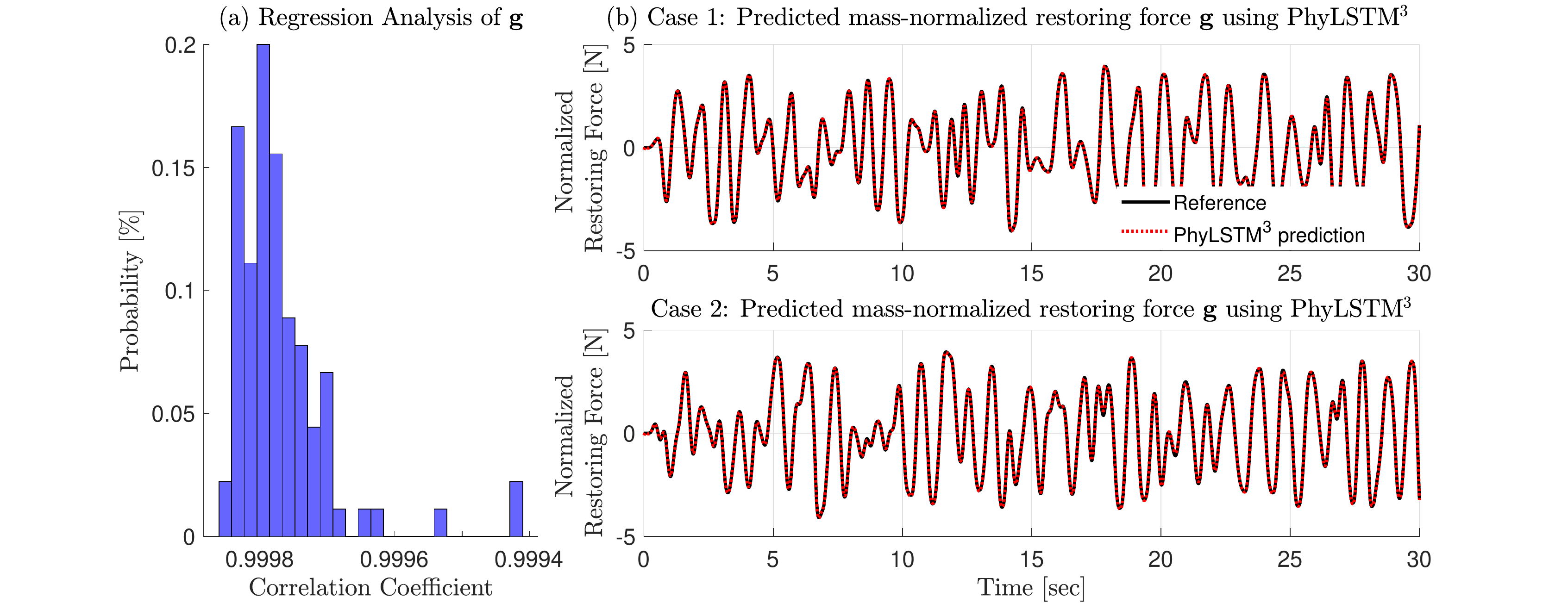}
	\caption{$\text{PhyLSTM}^3$-predicted mass-normalized restoring force: (a) regression analysis; and (b) predicted time histories.}
	\label{fig:g_BW_3LSTM}
\end{figure}

\begin{figure}[t!]
	\centering
	\includegraphics[width=1\linewidth]{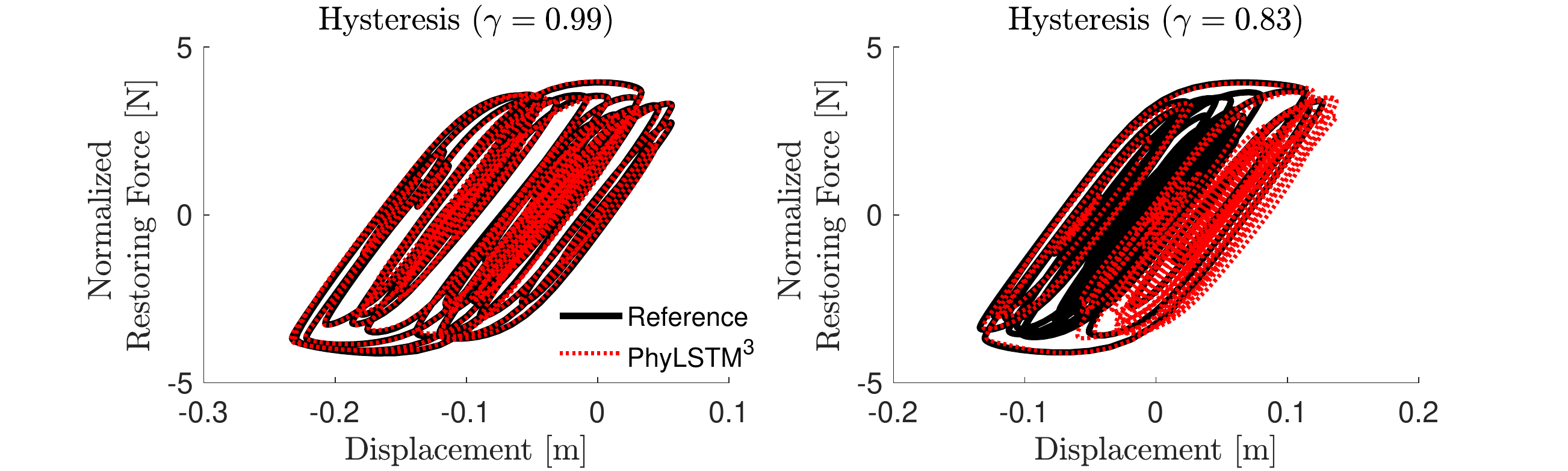}
	\caption{Examples of predicted hysteresis curves of nonlinear restoring force versus displacement using the proposed $\text{PhyLSTM}^3$.}
	\label{fig:hysteresis_3LSTM}
\end{figure}

\begin{figure}[t!]
	\centering
	\includegraphics[width=1\linewidth]{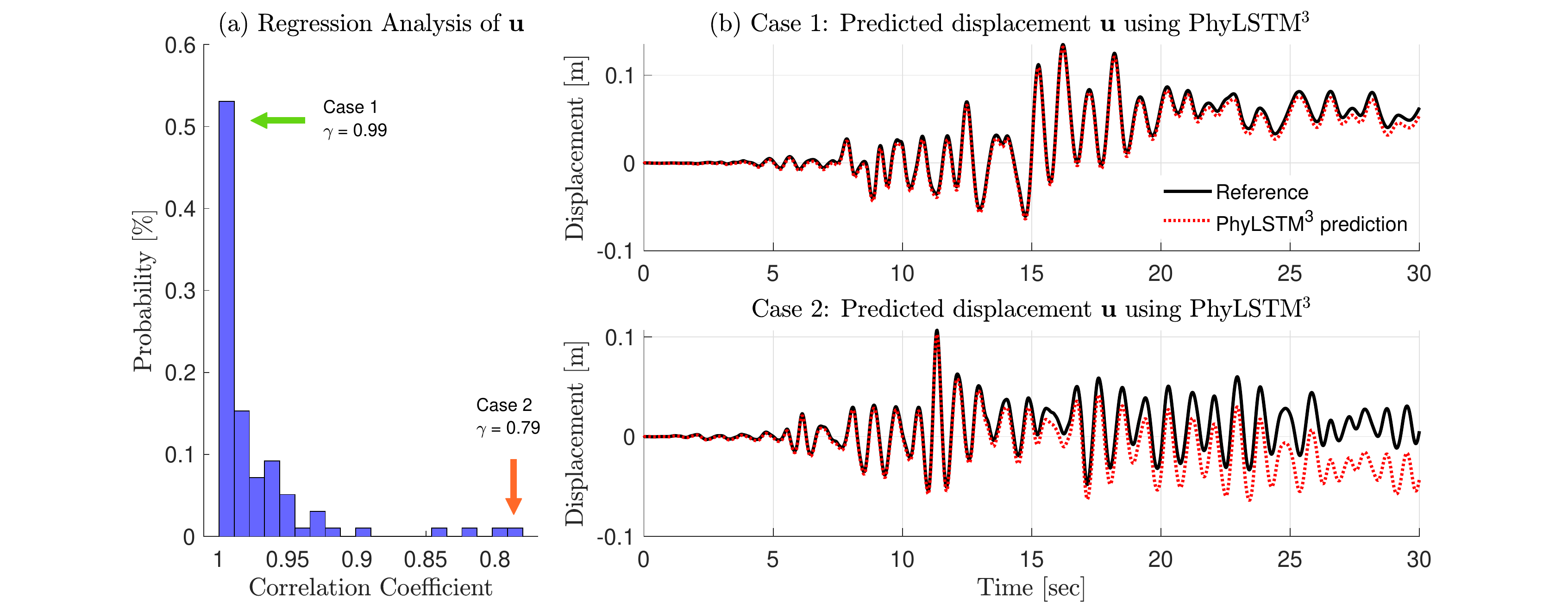}
	\caption{Predicted displacements of the SDOF Bouc-Wen model under unseen earthquake records using the $\text{PhyLSTM}^3$ metamodel trained by BLWN excitation data.}
	\label{fig:u_3LSTM_EQ}
\end{figure}

Figure \ref{fig:u_2LSTM_3LSTM} summarizes the performance of both $\text{PhyLSTM}^2$ and $\text{PhyLSTM}^3$ for prediction of nonlinear displacement time histories of the SDOF Bouc-Wen model under unseen BLWN excitations. Comparing the regression analysis shown in Figure \ref{fig:u_2LSTM_3LSTM}(a) and Figure \ref{fig:u_2LSTM_3LSTM}(c) for $\text{PhyLSTM}^2$ and $\text{PhyLSTM}^3$ respectively, it can be clearly seen that $\text{PhyLSTM}^3$ ensures a larger probability of correlation coefficients close to one, demonstrating a better prediction performance. Besides, the accuracy for the worst scenario using $\text{PhyLSTM}^3$ ($\gamma=0.77$) is much higher in contrast to $\text{PhyLSTM}^2$ ($\gamma=0.19$), indicating that $\text{PhyLSTM}^3$ is a more robust and stable approach for nonlinear rate-dependent hysteresis modeling. Figure \ref{fig:u_2LSTM_3LSTM}(b) shows two examples of predicted displacement time histories using $\text{PhyLSTM}^2$ and $\text{PhyLSTM}^3$ with the corresponding correlation coefficients of $\gamma=0.85$ and $\gamma=0.99$ for Case 1 and $\gamma=0.19$ and $\gamma=0.77$ for Case 2. The mass-normalized restoring force $g$ can be perfectly predicted (with $\gamma\approx1$) using the proposed $\text{PhyLSTM}^3$ as shown in Figure \ref{fig:g_BW_3LSTM} even though no measurement is available in training. The hysteresis of this nonlinear system can also be well estimated by the trained $\text{PhyLSTM}^3$ metamodel as depicted in Figure \ref{fig:hysteresis_3LSTM} which presents two examples of $u$-$g$ curves (e.g., predicted displacement v.s. predicted restoring force). To further test the robustness of the proposed approach, the $\text{PhyLSTM}^3$ metamodel trained by BLWN excitation data is employed to predict structural responses subjected to the suite of 97 ground motions used in the previous example. Figure \ref{fig:u_3LSTM_EQ}(a) summarizes the overall prediction performance over all 97 records using $\text{PhyLSTM}^3$, as a result, with the majority (e.g., $>$ 95\%) of correlation coefficients greater than 0.9. Figure \ref{fig:u_3LSTM_EQ}(b) shows two example time histories of predicted structural displacement with $\gamma=0.99$ and $\gamma=0.79$ (e.g., the worst scenario). In general, this clearly demonstrates the robustness of $\text{PhyLSTM}^3$ in metamodeling of nonlinear hysteretic system.

\section{Conclusions}\label{sec:conc}
This paper presents a novel physics-informed deep learning paradigm for metamodeling of nonlinear structural systems with showcase of predicting nonlinear structural seismic responses. In particular, two architectures of physics-informed multi-LSTM networks (e.g., $\text{PhyLSTM}^2$ and $\text{PhyLSTM}^3$) are presented for representation learning of sequence-to-sequence features from limited data enhanced by available physics. 
The laws of physics are taken as extra constraints, encoded in the network architecture, and embedded in the overall loss function to enforce the model training in a feasible solution space. In such way, the trained metamodel can accurately capture structural dynamics even with very scarce training/validation data. Another distinction of the proposed networks is that they can accurately model non-observable, latent nonlinear state variables (e.g., hysteretic parameter or nonlinear restoring force), where measurement is unavailable. 
The performance of $\text{PhyLSTM}^2$ and $\text{PhyLSTM}^3$ is demonstrated through two numerical examples (e.g., a 3-story MRF structure and a SDOF Bouc-Wen model). Numerical results illustrate that the physics-informed multi-LSTM models outperform the classical non-physics-guided data-driven LSTM network in terms of robustness and prediction accuracy. For nonlinear systems with rate-independent hysteresis, $\text{PhyLSTM}^2$ is more capable of modeling the latent nonlinearity given its parsimonious architecture compared with \text{PhyLSTM$^3$}; however, for the system with rate-dependent hysteresis, \text{PhyLSTM$^3$} is more powerful and produces much more accurate prediction thanks to its explicit modeling of the rate-dependent hysteresis using a differential equation. In general, the proposed $\text{PhyLSTM}^2$ and $\text{PhyLSTM}^3$ metamodels possess salient features that include (1) clear interpretability with physics meaning, (2) superior generalizability with robust inference, and (3) excellent capability of dealing with less rich data.
It turns out that the embedded physics can provide constraints to the network outputs, alleviate overfitting issues, reduce the need of big training datasets, and thus improve the robustness of the trained model for more reliable prediction. Though the proposed metamodeling approaches are presented in the context of structural seismic response prediction, they can be easily extended to develop metamodels for other types of structural systems, where the physics-informed multi-LSTM network architectures should be adapted by changing the physics part as needed.


\bibliographystyle{elsarticle-num}
\bibliography{refs}

\end{document}